\documentclass{article}
\usepackage{jheppub}

\usepackage{color}
\usepackage{amsmath,physics,amsfonts}
\usepackage[caption=false]{subfig}

\newcommand{\roughly}[1]{\mathrel{\raise.3ex\hbox{$#1$\kern-0.85em
\lower1ex\hbox{$\sim$}}}}

\def\exd{{\hbox{d}}}

\def\Sch{Schr\"odinger}

\def\ba{\begin{eqnarray}}
\def\ea{\end{eqnarray}}
\def\be{\begin{equation}}
\def\ee{\end{equation}}

\def\bfx{{\bf x}}

\def\ssA{{\scriptscriptstyle A}}
\def\ssB{{\scriptscriptstyle B}}

\def\ssF{{\scriptscriptstyle F}}
\def\ssE{{\scriptscriptstyle E}}
\def\ssI{{\scriptscriptstyle I}}

\def\ssN{{\scriptscriptstyle N}}

\def\ssR{{\scriptscriptstyle R}}

\def\cE{\mathcal{E}}
\def\cF{\mathcal{F}}
\def\cG{\mathcal{G}}

\def\cJ{\mathcal{J}}
\def\cK{\mathcal{K}}
\def\cL{\mathcal{L}}
\def\cN{\mathcal{N}}
\def\cO{\mathcal{O}}
\def\cR{\mathcal{R}}

\def\cT{\mathcal{T}}

\def\cW{\mathcal{W}}

\def\UV{{\scriptscriptstyle UV}}

\def\nn{\nonumber}

\def\({\left(}
\def\){\right)}

\def\pref#1{(\ref{#1})}

\def\tpsi{\tilde{\psi}}

\usepackage{graphicx}% Include figure files
\usepackage{dcolumn}% Align table columns on decimal point
\usepackage{bm}% bold math
\usepackage{color}

\title{Effective Field Theory of Black Hole Echoes}

\author[a,b]{C.P.~Burgess,}
\author[a,b]{Ryan Plestid}
\author[a,b]{and Markus Rummel}
\affiliation[a]{Physics \& Astronomy, McMaster University, Hamilton, ON, Canada, L8S 4M1}
\affiliation[b]{Perimeter Institute for Theoretical Physics, Waterloo, Ontario N2L 2Y5, Canada }
\emailAdd{cburgess@perimeterinstitute.ca} 
\emailAdd{plestird@mcmaster.ca}
\emailAdd{rummelm@mcmaster.ca}

\date{}

\abstract {Gravitational wave `echoes' during black-hole merging events have been advocated as possible signals of modifications to gravity in the strong-field (but semiclassical) regime. In these proposals the observable effect comes entirely from the appearance of nonzero reflection probability at the horizon, which vanishes for a standard black hole. We show how to apply EFT reasoning to these arguments, using and extending earlier work for localized systems that relates choices of boundary condition to the action for the physics responsible for these boundary conditions. EFT reasoning applied to this action argues that linear `Robin' boundary conditions dominate at low energies, and we determine the relationship between the corresponding effective coupling (whose value is the one relevant low-energy prediction of particular modifications to General Relativity for these systems) and the phenomenologically measurable near-horizon reflection coefficient. Because this connection involves only near-horizon physics it is comparatively simple to establish, and we do so for perturbations in both the Schwarzschild geometry (which is the one most often studied theoretically) and the Kerr geometry (which is the one of observational interest for post-merger ring down). In passing we identify the renormalization-group evolution of the effective couplings as a function of a regularization distance from the horizon, that enforces how physics does not depend on the precise position where the boundary conditions are imposed. We show that the perfect-absorber/perfect-emitter boundary conditions of General Relativity correspond to the only fixed points of this evolution. Nontrivial running of all other RG evolution reflects how modifications to gravity necessarily introduce new physics near the horizon.}

\begin{document}
%\preprint{IGC-17/8-2}

\maketitle
\section{Introduction and summary}

Orbiting systems famously decay because of the universal propensity of accelerating particles to emit radiation. A century ago the puzzle this raised about the stability of atoms when applied to electrically charged particles ultimately led to the development of quantum mechanics. More recently observations of gravitational waves emitted by mutually orbiting black holes and neutron stars \cite{LIGO} has caught them in the act of coalescing after complete orbital decay. 

Such measurements open up completely new ways to test how gravity behaves in the strong-field regime \cite{StrongFieldGRTests}. The bracing new breezes that this newly open observational window introduce promise to air out the musty room of theoretical speculation about the fundamental nature of black holes.

At first sight obtaining useful constraints from black-hole mergers might not seem likely given that no-hair theorems \cite{NoHair} suggest that field configurations exterior to black holes should robustly look like those of General Relativity (GR), even in the presence of additional light fields. Although these theorems are only as good as their assumptions, which might not be valid in any particular ultraviolet (UV) modification at high energies, they do seem to hold when the theory's low-energy limit contains finite numbers of low-spin fields (as many do). 

A variety of theoretical proposals \cite{UVproposals} argue that UV modifications might nonetheless fundamentally change black-hole structure. These alternatives include (but are not restricted to) the proposal that enormous numbers of new degrees of freedom appear where a black-hole horizon would naively be expected to form \cite{ComplexBHs}. The result is to replace black holes by hypothetical new black objects, with the event horizon acquiring a local physical significance not present in General Relativity. 

But how do these proposals get around the no-hair theorems to modify measurements by observers (like us) far from the black object? In general this is poorly understood, but one way to modify exterior solutions is by changing the near-horizon boundary conditions satisfied by external low-energy fields. Because any such a change involves moving away from the usual perfect-infall boundary conditions of black hole physics, it necessarily mixes in an overlap with an outgoing wave near the horizon. This then suggests searching for observable consequences of such outgoing solutions far from the black object to which the alternative boundary conditions might give rise. 

In particular this line of reasoning suggests that any initial burst of gravitational waves during a black-hole merger could be accompanied by a series of echoes as originally discussed in \cite{EchoLit} and from a more phenomenological perspective in \cite{EchoLitPheno,Nakano,Bueno:2017hyj,Correia:2018apm,PaniTesta,Mark:2017dnq}. The echo arises because the new boundary condition allows any infalling wave initiated outside the would-be horizon to be partially reflected at the horizon rather than falling completely in. This reflected wave can either escape to infinity or be reflected back towards the black object on its way out. If reflected, the process can repeat, leading to an ever diminishing series of echoes with a transfer function for each mode of the form \cite{Mark:2017dnq}

\begin{equation}
\mathcal{K} = \cT_\text{BH} \cR \sum_{i=0}^\infty (\cR_\text{BH} \cR)^i = \frac{ \mathcal{T}_\text{BH} \mathcal{R}}{1-\mathcal{R}_\text{BH}\mathcal{R}} \,,
\label{eq:transfer-functionintro}
\end{equation}
where $|\cR|^2$ represents the reflection probability for an inward-directed wave at the horizon, while $|\cR_\text{BH}|^2$ and $|\cT_\text{BH}|^2$ are respectively the probabilities of reflection and transmission for an outbound wave climbing through the gravitational field towards infinity. In this language the standard black-hole wave amplitude is obtained in the limit $\cR \to 0$, for which $\cK \to 0$.

Several possible echo signals have been computed this way, often through the imposition of ad-hoc boundary conditions (like Dirichlet conditions) at the position of the would-be horizon. Dirichlet conditions can be regarded as the most optimistic case, for which a wave is entirely reflected at the horizon rather than being only partly reflected there. 

Should such echoes be observed\footnote{There are even claims for low-significance evidence that they have been seen in the LIGO data \cite{Niayesh}. However, the detection/significance is challenged in \cite{NoEchoDetection}, see also \cite{NoNoEchoDetection}.} during the ring-down that follows an orbital in-spiral and merger, their two main observable features can be measured. The first of these is the time delay, $\Delta t = t_{i+1} - t_i$, between the `$i$th' echo, $S_i = S(t_i)$, and its successor, where $S(t)$ is a measure of the radiated wave amplitude. The second observable is the amount of damping between one echo and the next: $D = |S_{i+1}/S_i|$. 

The first of these, $\Delta t$, measures the phase of the combination $\cR \cR_\text{BH}$ while the second, $D$, measures its modulus. Given that $\cR_\text{BH}$ can be computed without reference to UV physics from standard properties of black hole geometries, measurements of both $\Delta t$ and $D$ allow a reconstruction of both the modulus and phase of $\cR$. 

\subsection*{This paper} 

So what does this paper add to the discussion? We add several new ingredients:

\medskip\noindent$\bullet$ {\it Extension to Kerr geometry}:

The first ingredient we add is an extension of the discussion of modified boundary conditions and new physics to the Kerr geometry. Previous treatments of black-hole boundary conditions are largely performed for the Schwarzschild metric,\footnote{See however \cite{Nakano,Bueno:2017hyj,Maggio:2017ivp}.} but the extension to Kerr is important because this is the geometry most relevant to post-merger ring-downs. 

\medskip\noindent$\bullet$ {\it Extension to spins one and two}:

Most explicit discussions of new near-horizon physics are couched for simplicity in terms of scalar fields. A second ingredient we add here is to extend the discussion to include spins one and two.\footnote{See \cite{PaniTesta,Maggio:2018ivz} for previous work on spin one and two in Schwarzschild.} We are able to do so because the main influence of new physics occurs in the near-horizon regime for which higher-spin equations share their key properties with scalars. 

\medskip\noindent$\bullet$ {\it Robust low-energy boundary condition}:

There is an enormous freedom in possible boundary conditions that can be imposed at the horizon. For example, although linear equations of motion may accurately describe the propagation of modes far from the horizon, there is nothing that forbids new physics (such as fuzzballs or firewalls) from being highly non-linear; thus, boundary conditions need not be linear (i.e.\ Dirichlet, Neumann or Robin), and can assume the form $\partial_r \psi=g(\psi)$ for any function $g(\psi)$. Our contribution is to provide a physical criterion for how to identify which near-horizon boundary conditions are most likely to be robustly relevant at low energies for many types of UV modifications to GR. We are able to discriminate amongst boundary conditions using low energies as a the criterion by introducing a near-horizon boundary component to the action used to describe the influence of new physics on the low-energy effective field theory (EFT) outside the black object. The required boundary conditions are then related to the boundary action by the condition that the total (bulk plus boundary) action be stationary with respect to arbitrary variations of the fields on the boundary.

Expressing boundary conditions in terms of boundary contributions to the effective action is useful because we know how to rank interactions in an effective action according to their low-energy impact. Because EFTs are dominated at low energies by interactions involving the fewest fields and derivatives it is generic (given some symmetries, like $\psi \to - \psi$) that the lowest-dimension interaction on the boundary is proportional to $\psi^2$. This suggests that linear `Robin' boundary condition of the form 
\be \label{RobinBCs}
   \partial_r \psi = \lambda \, \psi
\ee
should very generically dominate in the near-horizon region at low energies for a broad class of new physics \cite{Robin, PPEFT1}.

\medskip\noindent$\bullet$ {\it RG-invariant relations between couplings and observables}:

Once boundary conditions begin to differ from the standard perfect-infall conditions usually used for black holes it becomes generic that many quantities diverge at the position of the horizon. We regulate these divergences by imposing the boundary condition \pref{RobinBCs} a small distance away from the horizon, at $r = r_+ + \epsilon$ where $r_+$ is the Kerr outer horizon and $0 < \epsilon \ll r_+$. Because there is a great freedom in the values of $\epsilon$ that can be chosen when doing so, nothing physical can depend on the precise value of $\epsilon$ used. This happens in practice because any explicit dependence on $\epsilon$ in any calculation is precisely cancelled by an implicit dependence $\lambda(\epsilon)$ in the boundary condition \pref{RobinBCs}. 

We determine the renormalization-group (RG) equation that determines how $\lambda$ depends on $\epsilon$ in order to ensure the $\epsilon$-independence of physical quantities. We do so fairly generally for spherically symmetric new physics for Schwarzschild, but for simplicity only deal explicitly with boundary conditions that do not mix modes for Kerr geometries.
Using these flow equations we show that the standard perfect-infall (or, for white holes, perfect-outflow) boundary conditions correspond to the only two RG fixed points. Being a fixed point coincides with standard black-hole lore in GR: the horizon is not locally a special place, and so the choice of effective couplings corresponding to GR should be explicitly $\epsilon$-independent. As a corollary deviations from standard-GR black-hole physics always introduce nontrivial RG flow to these effective couplings, indicating how the horizon always develops a local physical significance for such modifications.

We argue that the RG flow equations imply that any RG-invariant quantity, including any physical observable, can be efficiently characterized in terms of two RG-invariant real parameters: a dimensionless quantity, $z_\star$, and a length scale $\epsilon_\star$. The scale $\epsilon_\star$ should not be confused with the scale $\epsilon_0$ where the boundary condition \pref{RobinBCs}, say, is imposed. $\epsilon_\star$ is instead an RG-invariant way to label the particular RG-flow line, $\lambda(\epsilon)$ on which the point $\lambda(\epsilon_0) = \lambda_0$ is located. In particular, it can happen that $\epsilon_\star$ is much greater or smaller than $\epsilon_0$.

We also compute explicitly how the main observable --- the near-horizon reflection coefficient $\cR$  --- depends on the RG-invariant parameters $z_\star$ and $\epsilon_\star$, as well as how $\cR$ depends on $\lambda(\epsilon_0)$ and $\epsilon_0$. The main results are given by eqs.~\pref{Rvsxilambda} and \pref{RforKerr} for Kerr geometries (which include Schwarzschild geometries as a special case). We find the implications of these formulae for echo delays, $\Delta t$, and echo damping, $D$.

%\pagebreak
\medskip\noindent$\bullet$ {\it Relation to the Schr\"odinger problem with inverse-square potential}:

Finally, much of the previous discussion is made relatively easy by mapping the radial part of the Teukolsky equation~\cite{Teukolsky:1973ha} into a radial Schr\"odinger equation. By doing so we show how the above black-object story directly maps onto arguments coming from point-particle effective field theories (PPEFTs), used elsewhere to describe the influence of small central sources (like nuclei) on much larger orbits (like those for atomic electrons) described using the Schr\"odinger \cite{PPEFT1}, Klein-Gordon \cite{PPEFT2} and Dirac equation \cite{PPEFT3}. (For related discussions of inverse-square potentials and conformal properties in the near-horizon limit see \cite{carlos, steve}.)

In particular, under this mapping the near-horizon mode evolution for the Kerr metric maps onto the scale-invariant problem of a Schr\"odinger field interacting with an inverse-square potential \cite{OtherSchInvSq}, with the origin of the potential mapping onto the position of the horizon. Probability loss through the horizon then corresponds to there also being probability loss at the origin in the \Sch\ problem,\footnote{For super-radiant modes in a Kerr spacetime the origin instead turns out to be source of probability in the \Sch\ analog problem.} corresponding to non-self-adjoint extensions to the inverse-square problem. Similar central probability loss for \Sch\ fields with inverse-square potentials can also occur, one example of which is given by trapped polarizable atoms interacting with a hot wire, since this is also described by an inverse-square potential plus a probability sink at the origin. We find the black-hole examples closely follow the PPEFT treatment of this hot-wire system \cite{PPEFT4}.  

These results are presented in the following way.  First, \S\ref{sec:ModeFn} sets up the tools needed for the rest of the paper. This starts, in \S\ref{ssec:KerrKG}, with a summary of required results about fields propagating on a Kerr geometry. This section includes also a discussion of the map of the radial mode equation into an equivalent radial \Sch\ equation in $d$ spatial dimensions.

The guts of the discussion is contained in \S\ref{sec:NearHorizonBC}. This first defines, in \S\ref{sec:ReflecTrans}, the near-horizon reflection coefficient $\cR$ that is the main quantity of practical interest. This section shows how this parameter and its implications for cross-horizon probability flow can all be identified using only asymptotic near-horizon information. Next \S\ref{ssec:BCEFTs} describes how to relate boundary conditions at a radius $r = r_+ + \epsilon$ to a boundary action, and argues why dimensional arguments for effective couplings in this action imply linear Robin conditions should dominate at low energies. This section also shows how this boundary story maps in the inverse-square \Sch\ version of the problem onto probability loss at the origin. Finally, \S\ref{ssec:RGFlow} sets up the RG flow equations that ensure physics doesn't depend on a particular value of the arbitrary regularization parameter $\epsilon$, and computes $\cR$ in terms of the RG-invariant parameters $z_\star$ and $\epsilon_\star$.

Finally, \S\ref{sec:Pheno} makes the connection between the formalism of earlier sections and the phenomena accessible for measurements of gravitational wave echoes. In particular it connects echo time-delay and damping to $\cR$ and uses this to probe how these depend on the two RG-invariant measures of new physics, $z_\star$ and $\epsilon_\star$. The main results are briefly summarized in \S\ref{sec:Concl}.

\section{Mode functions}
\label{sec:ModeFn}

This section establishes the `bulk' physics of interest describing a linearized fluctuation in a black-hole background spacetime. In particular we follow standard practice by mapping the equation governing radial motion into a Schr\"odinger-like form, for which considerable intuition applies. This is mostly bread-and-butter stuff, so aficionados should feel free to skip ahead to the discussion of near-horizon boundary conditions.

\subsection{Background metric}

Echoes are observationally sought in the aftermath of a merger with the black object generically rotating, so consider first the Kerr metric in Boyer-Lindquist coordinates,
\be \label{Kerr}
  \exd s^2 = - \frac{\Delta}{\rho^2} \Bigl[ \exd t - a \sin^2 \theta \, \exd \phi \Bigr]^2 + \frac{\sin^2\theta}{\rho^2} \Bigl[ (r^2+a^2) \exd \phi - a\, \exd t \Bigr]^2+ \frac{\rho^2}{\Delta} \, \exd r^2 + \rho^2 \exd \theta^2   \,,
\ee
where the functions $\rho(r,\theta)$ and $\Delta(r)$ are given explicitly by
\be
 \Delta := r^2 - r_sr + a^2 \,,
\ee
and
\be
  \rho^2 := r^2 + a^2 \cos^2 \theta \,,
\ee
while $a = J/M$ measures the black object's angular momentum while $r_s = 2GM$ parameterizes its mass.

According to this metric the proper time of a hovering observer (at fixed $\{r, \theta, \phi\}$) is given by
\be
  \exd \tau^2 =  \left( 1 - \frac{r_s r}{\rho^2} \right) \exd t^2 \,,
\ee 
which shows that such observers are only possible outside the ergosphere, $r > r_\ssE(\theta)$, with
\be \label{ergo}
  r_\ssE(\theta) > \frac12 \Bigl[ r_s + \sqrt{r_s^2 - 4a^2\cos^2 \theta}\Bigr] \,,
\ee
and for radii smaller than this all timelike observers are necessarily dragged in the direction of the black hole's rotation. The (outer) horizon of this geometry is located at $r = r_+$ with
\begin{equation}
 r_+ = \frac{r_s + \sqrt{r_s^2-4a^2}}{2}\,.
\end{equation}
This is real for $|a| \le GM = r_s/2$ and $0 \le r_+ \le r_s$ throughout this range. The ergosphere and horizon coincide at the poles, where $\cos^2\theta = 1$, but for other angles $r_\ssE > r_+$.

\subsection{Perturbations in a Kerr background}
\label{ssec:KerrKG}

Our interest external to the black object is in linearized fields propagating on a background metric, which we take to be well-approximated by the Kerr metric. For linearized fields, it was first shown by Teukolsky~\cite{Teukolsky:1973ha} that a second order differential equation can be formulated:
\begin{align}\label{Teukolsky}
 \begin{aligned}
  &\left[\frac{(r^2+a^2)^2}{\Delta} - a^2 \sin^2 \theta \right] \frac{\partial^2 \psi}{\partial t^2} + \frac{2 r_s r a}{\Delta} \frac{\partial^2 \psi}{\partial t \partial \phi} + \left(\frac{a^2}{\Delta} - \frac{1}{\sin^2 \theta} \right) \frac{\partial^2 \psi}{\partial \phi^2} - \Delta^{-s} \frac{\partial}{\partial r} \left(\Delta^{s+1} \frac{\partial \psi}{\partial r} \right)\\ 
  &- \frac{1}{\sin \theta} \frac{\partial}{\partial \theta} \left(\sin \theta \frac{\partial \psi}{\partial \theta} \right) -s \left[\frac{a(2r-r_s)}{\Delta} + 2i\frac{\cos \theta}{\sin^2 \theta} \right] \frac{\partial \psi}{\partial \phi} - s \left[\frac{r_s(r^2-a^2)}{\Delta} -2r - 2ia \cos \theta \right] \frac{\partial \psi}{\partial t}\\
  &+ \left(s^2 \cos^2 \theta - s \right) \psi = 0\,,
 \end{aligned}
\end{align}
where $s$ refers to the propagation of scalars $(s=0)$, Dirac fermions $(s= \pm 1/2)$, electromagnetic $(s= \pm 1)$, and gravitational $(s= \pm 2)$ fields (all massless). We list the definition of the field $\psi$ for spin zero, one and two in Table~\ref{psiNP_table} (for more details see~\cite{Teukolsky:1973ha}). 
\begin{table}[ht!]
\centering
  \begin{tabular}{c||c|c|c|c|c}
   $s$ & $0$ & $1$ & $-1$ & $2$ & $-2$\\
  \hline
  $\psi$ & $\phi$ & $F_{\mu\nu}l^\mu m^\nu$ & $\tilde \rho^{-2} F_{\mu\nu}m^{*\mu}n^\nu$ & $-C_{\alpha\beta\gamma\delta}l^\alpha m^\beta l^\gamma m^\delta$ & $-\tilde \rho^{-4} C_{\alpha\beta\gamma\delta}n^\alpha m^{*\beta} n^\gamma m^{* \delta}$\\
  \end{tabular}
  \caption{The field $\psi$ for different spin weight $s$ in the Teukolsky equation~\eqref{Teukolsky}. $C_{\alpha\beta\gamma\delta}$ is the Weyl tensor and $\tilde \rho = -(r- i a \cos \theta)$. The quantities $l^\mu = \left[(r^2+a^2)/\rho^2,1,0,a/\rho^2 \right]$, $n^\mu = \left[r^2+a^2,-\rho^2,0,a \right]/(r^2 + a^2 \cos^2 \theta)$ and $m^\mu = \left[i a \sin \theta, 0, 1, i/\sin\theta \right]/2^{1/2}(r+ia \cos \theta)$ are known as Kinnersley's null tetrad~\cite{Kinnersley:1969zz}.}
  \label{psiNP_table}
\end{table}

For the Klein-Gordon field $(s=0)$ \eqref{Teukolsky} is simply $\Box \psi = \frac{1}{\sqrt{-g}}\partial_\mu \left( \sqrt{-g} g^{\mu\nu}  \partial_\nu \psi \right)$ which can be derived from the action
\begin{equation}\label{SB}
 S_\ssB = - \frac12 \int \exd^4 x \sqrt{-g}  g^{\mu\nu} \partial_\mu \psi \,\partial_\nu \psi \,.
\end{equation}

\subsection{Mode decomposition}

Equation~\eqref{Teukolsky} separates in Boyer-Lindquist coordinates for $r>r_+ + \epsilon$ with the {\it ansatz} 
\be \label{psisep}
  \psi(t,r,\theta,\phi) = R(r) \,S(\theta) \,e^{-i\omega t} \,e^{i m \phi} \,, 
\ee
leading to separated ordinary differential equations for $S(\theta)$ and $R(r)$. Once boundary conditions are specified at the horizon and at spatial infinity the frequency $\omega$ in general can be complex (as is the case for quasi-normal modes (QNMs). Alternatively, for scattering states $\omega$ can be chosen to be real with boundary conditions imposed only at the horizon, with scattering amplitudes read off from the asymptotic form at infinity. 

The angular equation is
\begin{equation}
 \frac{1}{\sin \theta} \frac{\exd}{\exd \theta} \left(\sin \theta\, \frac{\exd S}{\exd \theta}  \right) + \left[ a^2 \omega^2 \cos^2 \theta - \frac{m^2}{\sin^2 \theta} - 2 a \omega s \cos \theta - \frac{2ms \cos \theta}{\sin^2 \theta} - s^2 \cot^2 \theta + s  +A_{slm} \right] S  = 0\,,
\end{equation}
This equation is solved by oblate spheroidal harmonic functions, $S_{lm}(i \omega,\cos \theta)$ with eigenvalues $A_{lm}$. Here the mode labels are $m$ and $l$ and periodicity under $2\pi$ shifts of $\phi$ together with boundedness at $\cos\theta = \pm 1$ imply both are integers, satisfying $|m|\leq l$, for details see~\cite{BertiCardoso}. It is useful to denote the entire angular modes by $Z_{lm}(\theta,\phi) = S_{lm}(\theta) \, e^{im\phi}$, and in the special case of the Schwarzschild limit ($a \to 0$) for a scalar field $s=0$ these angular mode functions become the usual spherical harmonics: $Z_{lm}(\theta,\phi) \to Y_{lm}(\theta,\phi)$ and $A_{slm} \to l(l+1)$.

The radial equation similarly is
\begin{equation}\label{Req}
%  \Delta \frac{d}{dr} \left(\Delta \frac{d R(r)}{d r} \right) + \tilde V(r) R(r) = 0\,,
 \frac{\exd^2 R}{\exd r^2} + \frac{(2r-r_s)(s+1)}{\Delta} \left( \frac{\exd R}{\exd r} \right) - \frac{W(r)}{\Delta^2} \;R = 0\,,
\end{equation}
with
\begin{align}\label{Wdef}
 \begin{aligned}
  W(r) := -\left(r^2+a^2\right)^2 \omega^2 &+ 2 a r_s r \omega m - a^2 m^2 - is \left[(2r-r_s) ma - (r^2-a^2) r_s \omega \right]\\
  &+ \left( a^2 \omega^2  + A_{slm}  - 2i s r \omega\right)\Delta\,.
 \end{aligned}
\end{align}

\subsection{\Sch\ form}

In order to make contact with reasoning in other systems, we now rewrite the radial equation, \eqref{Req}, in \Sch\ form. That means we redefine the radial function 
\begin{equation}
  \label{eq:radial-ansatz}
 R(r) =: F_d(r)\,\chi_d(r)\,,
\end{equation}
in such a way as to arrange that the derivative terms for the new variable combine into the particular combination 
\be \label{chiderivs}
  \frac1{(r-r_+)^{d-1}} \, \frac{\exd}{\exd r} \left[ (r-r_+)^{d-1}\, \frac{\exd \chi_d}{\exd r} \right] = \chi''_d + \left( \frac{d-1}{x} \right) \chi'_d \,,
\ee
where primes denote differentiation with respect to $x := r-r_+$. Eq.~\pref{chiderivs} is chosen to agree with the radial part of the flat-space laplacian, $\nabla^2 \chi_d$, in $d$ spatial dimensions, with $x$ regarded as the radial coordinate.\footnote{The two most natural choices are $d=1$ (because the horizon is a codimension-one surface) or $d=3$ (if singularities at the horizon are compared with singularities at the origin for polar coordinates on flat space). In what follows we show how these two choices lead to \Sch\ equations with indistinguishable potentials.}

Substituting \pref{eq:radial-ansatz} into \pref{Req} shows that the function $F(r)$ required to effect this change must satisfy
\be \label{Feq}
 \frac{2F_d'}{F_d} + \frac{(2r -r_s)(s+1)}{\Delta}  =  \frac{d-1}{r-r_+} \,.
\end{equation}
With this choice \pref{Req} takes the \Sch-like form
\be \label{chieom}
 -\nabla^2 \chi_d + V_d(x) \, \chi_d = 0 \,,
\ee
with `potential' given by
\begin{equation}\label{VR}
 V_d = \frac{W}{\Delta^2} -  \frac{(2r-r_s)(s+1)}{\Delta} \frac{F_d'}{F_d} - \frac{F_d''}{F_d} \,.
\end{equation}

To determine $F_d(r)$ we solve eq.~\pref{Feq}, which using $\Delta'(r) = 2r - r_s$ can be written
\begin{equation}
 \frac{2F_d'}{F_d} + (s+1)\frac{\Delta'}{\Delta} = \frac{\exd}{\exd r} \Bigl[ \ln \Bigl( \Delta^{s+1} F_d^2 \Bigr) \Bigr] = \frac{d-1}{r-r_+}\,,
\end{equation}
and so the general solution is
\begin{equation} \label{GenFsoln}
 F_d = C_\ssF \frac{(r-r_+)^{(d-1)/2}}{\Delta^{(s+1)/2}}\,,
\end{equation}
with integration constant $C_\ssF$ (whose precise value does not matter in what follows). With this choice the \Sch\ potential, \pref{VR}, becomes
\begin{equation}\label{VR2}
 V_d = \frac{W+(s+1)\Delta}{\Delta^2} - \frac{1-s^2}{4} \left( \frac{\Delta'}{\Delta} \right)^2 - \frac{(d-1)(d-3)}{4(r-r_+)^2} = V_1  - \frac{(d-1)(d-3)}{4(r-r_+)^2} \,.
\end{equation}
Explicitly, the $d$-independent term evaluates to
\begin{align}\label{V1}
 \begin{aligned}
  V_1 = &-\, \frac{ \left(1-s^2\right) \left(r_s^2/4 - a^2\right) + \left(r^2+a^2\right)^2\omega^2 - 2 r_s r a \omega m +a^2 m^2 + i s \left[(2r-rs)m a - (r^2-a^2) r_s \omega \right]}{\Delta^2}\\ &\qquad+ \frac{a^2 \omega^2 + A_{slm} + s(s+1) -  2i s \omega r}{\Delta} \,.
 \end{aligned}
\end{align}

For large $r$ because $\Delta \to r^2$ this potential asymptotes to an $a$-independent constant, $V_1(r) \to -\omega^2$. In the opposite, near-horizon, limit $r - r_+ \to 0^+$ the potential instead diverges because $\Delta \to 0$. Using the asymptotic form
\begin{equation}\label{Deltaapprox}
 \Delta \simeq \sqrt{r_s^2 - 4 a^2} (r-r_+)\,,
\end{equation}
in this limit, in the near-horizon regime the potential becomes
\begin{equation}\label{V1approx}
 V_1(r-r_+) = -\,\frac{1-s^2}{4(r-r_+)^2}-\, \frac{ (\omega r_+ r_s -  a  m)^2 + is\left[m a\sqrt{r_s^2-4a^2} - (r_+^2-a^2) r_s \omega \right]}{(r_s^2-4a^2)(r-r_+)^2} +  \mathcal{O}\left(\frac{1}{r-r_+}\right)\,,
\end{equation} 
which uses the identity $r_+^2 + a^2 = r_+ r_s$. 

It is this limiting form that controls the near-horizon physics, and in particular how modes respond to any new physics localized near the horizon. Because this is an inverse-square potential the system's near-horizon behaviour naturally becomes scale-invariant and universal. As has been remarked elsewhere for the Schwarzschild black hole \cite{OtherSchInvSq}, this inverse-square potential turns out always to be attractive because $r_s^2 \ge 4 a^2$. Because this leading inverse-square interaction is attractive the solutions to the \Sch\ equation are particularly sensitive to the boundary conditions chosen as $r \to r_+$, as we now describe.

\subsection{Near-horizon behaviour}

To summarize, the \Sch\ version of the radial mode equation is given by pure radial solutions to eq.~\pref{chieom}, 
\be
\label{chi3eom}
 \chi_d'' + \left( \frac{d-1}{x} \right) \chi_d' - V_d(x) \chi_d = 0 \,,
\ee
where for $0 < x = r- r_+ \ll r_s$ the potential has the asymptotic form
\begin{equation}
 V_d(x) = -\, \frac{v_d}{x^2} - \frac{w}{x} - \tilde k^2 + \cO(x^1) \,,
\end{equation}
with \pref{V1approx} implying the leading coefficient,
\ba \label{vdcoeffs}
  v_d &=& v_1 + \frac{(d-1)(d-3)}{4} \nn\\
  \hbox{with} \quad
  v_1 &=& \frac{1-s^2}{4} + \frac{(\omega r_+ r_s -  a  m)^2+ is\left[m a\sqrt{r_s^2-4a^2} - (r_+^2-a^2) r_s \omega \right]}{r_s^2-4a^2}  \,,
\ea
depends only on the conserved mode labels $\omega$ and $m$. By contrast the label $l$ enter first through the subleading coefficients,
\begin{align}
  w = &\omega (r_s\omega + i s)+\frac{8 a^4 \omega ^2+2 a^2 \left(4 A_{slm}-2 m^2-7 r_s^2 \omega ^2-4 i r_s s \omega +2 s^2+4 s+2\right)+4 a m r_s^2 \omega}{2 \left(r_s^2-4 a^2\right)^{3/2}} \nn\\
  &-\frac{r_s^2 \left[2 A_{slm}-2 r_s^2 \omega ^2+s (2-2 i r_s \omega )+s^2+1\right]}{2 \left(r_s^2-4 a^2\right)^{3/2}} \,,\\
 \tilde k^2 = &\frac{12 a^4 \omega ^2-a^2 \left(4 A_{slm}-3 m^2+2 r_s^2 \omega ^2-6 i r_s s \omega +s^2+4 s+3\right)-3 a m r_s^2 \omega }{ \left(r_s^2-4 a^2\right)^2} \nn\\
 &+\frac{r_s^2 \left(4 A_{slm}+2 r_s^2 \omega ^2-6 i r_s s \omega +s^2+4 s+3\right)}{4 \left(r_s^2-4 a^2\right)^2}-\frac{8 a^2 \omega  (r_s \omega +i s)-2 a m (r_s \omega -i s)-r_s^2 \omega  (r_s \omega +3 i s)}{2 \left(r_s^2-4 a^2\right)^{3/2}}\,. \label{ktilde}
\end{align}

It is the coefficient $v_1$ that dictates the near-horizon asymptotic form for the solutions, which take the form $\chi^\pm(x) \propto x^{b_\pm}$ for $x \to 0$ where the differential equation implies $b^2 + (d-2)b +v_d = 0$ and so
\ba
 b_\pm &=& \frac12 \Bigl[ - (d-2) \pm \sqrt{(d-2)^2 - 4v_d} \Bigr] \nn\\
 &=& \frac12 \Bigl[ -(d-2) \pm \sqrt{1 - 4 v_1} \Bigr] \,.
\ea
Notice that because $F_d(x) \propto x^{(d-s-2)/2}$ for small $x$ this implies $R^\pm(x) = F_d(x) \chi^\pm(x) \propto x^{\nu_\pm}$ with powers
\be
 \nu_\pm =  \frac12\left( -s \pm \sqrt{1-4v_1} \right)\,,
\ee
independent of $d$, as they must be. 

It is convenient to define $i \xi = \frac12 \, \sqrt{1-4v_1}$ which with \pref{vdcoeffs} gives
\begin{equation}
 \xi = \left\{\frac{(\omega r_s r_+-am)^2 + is\left[m a\sqrt{r_s^2-4a^2} - (r_+^2-a^2) r_s \omega \right]}{r_s^2-4a^2} - \frac{s^2}{4} \right\}^{1/2}  \,.
 \label{eq:zeta-def}
\end{equation}
In general $\xi = \xi_\ssR + i \xi_\ssI$ is complex we always choose the principal branch of the complex square root in \eqref{eq:zeta-def} such that $\xi_\ssI > 0$.  In the special case $s=0$ \textit{and} $\omega$ real, $\xi$ is real\footnote{Strictly speaking $\omega$ is only real if one does not consider quasi-normal modes for which more generically $\xi$ becomes complex once $\omega$ is complex even in the case $s=0$.} and given by
\begin{equation}
 \xi = \frac{\omega r_s r_+-am}{\sqrt{r_s^2-4a^2}}  \,.
 \label{eq:zeta-defs=0}
\end{equation}
%
% Because $s_\pm = \pm i \xi$ is imaginary, 
A physical interpretation for the solutions, $\psi^\pm(x)$, is found by reintroducing the mode's $t$-dependence, leading to the near-horizon behaviour
\begin{equation}
 \psi^\pm \propto R^\pm(x) \, e^{-i\omega t} \simeq x^{-s/2}\,e^{-i (  \omega t \mp  \xi \ln x )} \qquad \hbox{(for $x \to 0$)}\,.
\end{equation}
This reveals the solution $\psi_-(x)$ to be the purely infalling wave and the solution $\psi_+(x)$ to be the purely outgoing wave in the asymptotic near-horizon limit.

\subsection{Scalar Schwarzschild limit}

It is useful to quote the special case $a = s = 0$ to make contact with results for the Klein-Gordon field in the Schwarzschild geometry. In the scalar Schwarzschild case we have $a = s =0$ and $r_+ = r_s$ and so the \Sch\ potential reduces to the well-known $m$-independent expression
\begin{equation} \label{V1Schw}
 V_{1}(r-r_s) \to - \; \frac{(\omega^2 r_s^2 + 1/4)}{(r-r_s)^2} + \mathcal{O}\left(\frac{1}{r-r_s}\right) \qquad \hbox{(as $a,s \to 0$)}\,,
\end{equation}
and the asymptotic coefficients $v_1$ and $w$ simplify to 
\be
  v_1 \to (\omega r_s)^2 + \frac14 \quad \hbox{and} \quad
  w \to - \frac{2l(l+1) +1}{2r_s} +2\omega^2 r_s \qquad\hbox{(as $a,s \to 0$)}\,.
\ee
Finally, in the Schwarzschild limit the parameter $\xi$ becomes
\begin{equation}
 \xi  \to \omega r_s \qquad \hbox{(as $a,s \to 0$)}\,.
\end{equation}

\section{Near-horizon boundary conditions}
\label{sec:NearHorizonBC}

For each mode the general solution external to the black object is a linear combination of the radial solutions found above, most conveniently written
\begin{equation}
 \psi  = \psi^- + \cR \,\psi^+ \,,
 \label{eq:gensoln}
\end{equation}
for complex constant $\cR$, where the mode $\psi^\pm(t,r,\theta,\phi)$ has $R^\pm(r)$ as its radial component. 

To the extent that new physics involves only $r < r_+ + \epsilon$ (for $0 < \epsilon \ll r_s$) it can only affect physics at $r \gg r_+$ through the boundary condition that it imposes on the modes as $r \to r_+$ from above, and its implications for the integration constant $\cR$. In particular, as shown above, the standard `perfectly infalling' black-hole boundary condition corresponds to the particular choice $\cR = 0$ for each mode, so any boundary condition leading to a value different from this is the footprint of new physics.

This section makes two points about the constant $\cR$. First, we compute explicitly how this constant is connected to the main observable parameter: the reflection probability at the horizon. Then it is shown how the boundary condition that sets the value of $\cR$ can be determined from an EFT that captures the low-energy limit of any new physics near the horizon. This is special case of a more general story in which the low-energy effective action describing the physics of a small source determines the boundary condition for fields near the source \cite{PPEFT1}.  

\begin{figure}
\centering
\includegraphics[width=0.4\textwidth]{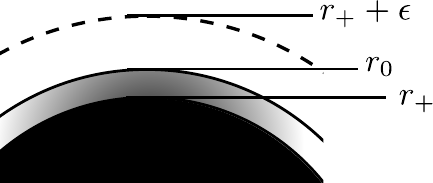}
\caption{The boundary condition is imposed at a radius $r_+ + \epsilon$ such that physical observables do not depend on the precise value of $\epsilon$. This is a cutoff independent way to describe new physics in the region $r_+ < r_0 < r_+ + \epsilon$ (gray) where $0 < \epsilon = r- r_+ \ll r_s$.}
\label{sketch_fig}
\end{figure}

Because new physics generically introduces singular behaviour at $r = r_+$ this boundary condition is actually imposed a short distance away, at a `stretched horizon' \cite{Stretched} with $0 < \epsilon = r- r_+ \ll r_s$, see also Figure~\ref{sketch_fig}. This section also computes the renormalization-group flow for EFT couplings that follows from the requirement that physical predictions do not depend on the arbitrary scale $\epsilon$.

\subsection{Klein-Gordon near-horizon reflection and transmission}
\label{sec:ReflecTrans}

The two basis mode functions of the Teukolsky equation have the asymptotic near-horizon form
\be \label{psiasymptotic}
  R^\pm(r) e^{-i\omega t} \simeq N^\pm \, \frac{1}{(r-r_+)^{s/2}}\, e^{- i \left[ \omega t \mp \xi \ln (r-r_+) \right]}  \qquad \hbox{(as $r \to r_+$)}
\ee
where the constants $N^\pm$ are calculable, in principle, by normalizing modes throughout the external region $r > r_+$. The determination of $N^\pm$ is a standard, though tedious, calculation to which new physics makes no contribution.

For comparison, the corresponding \Sch\ solution satisfies 
\be \label{chiasymptotic}
  \chi_d^\pm(r) e^{-i\omega t} \simeq \frac{\cN^\pm \, e^{-i [\omega t \mp \xi \ln(r-r_+)]} }{(r-r_+)^{ (d-2)/2}}  \qquad \hbox{(as $r \to r_+$)}\,,
\ee
where $\cN^\pm$ is similarly defined by demanding these modes be normalized in the region $r > r_+$ using the \Sch\ inner product. In the following, we discuss conserved quantities and the near-horizon probability flow. For simplicity, we restrict to the Klein-Gordon case $(s=0)$, as here the Teukolsky field is identical to the physical field $\psi = \phi$.

\subsubsection{Conserved quantities}

Reflection coefficients in the near-horizon regime can be computed in terms of the Klein-Gordon $(s=0)$ and \Sch\ conserved currents, so in this section we first briefly relate these to one another. 

On the \Sch\ side \Sch\ evolution, $i \partial_t \chi = (-\nabla^2+V) \chi$ implies $\partial_t(\chi^* \chi) + \nabla \cdot \cJ = 0$ for spatial probability flux current, 
\be
  \cJ_k :=  i \Bigl( \chi \,\partial_k \chi^*-  \chi^* \partial_k \chi  \Bigr)\,.
\ee
On the Klein-Gordon side the natural conserved current is the Wronskian
\be \label{Wronskian}
 \cW_\mu := i \Bigl(  \psi \,\partial_\mu \psi^* - \psi^* \partial_\mu \psi \Bigr) \,,
\ee
which the Klein-Gordon equation ensures satisfies $\partial_\mu (\sqrt{-g}\; \cW^\mu ) = 0$. 

A second conserved vector on the Klein-Gordon side is the energy current --- defined as\footnote{The negative sign here ensures that it is $E^t = + T_{tt}$ is the energy density at infinity where $g_{tt} \to -1$.} $E_\mu := -T_{\mu\nu} K^\nu$, where $K^\mu \partial_\mu = \partial_t$ is the metric's timelike Killing vector --- since this is also conserved in the sense that $\partial_\mu (\sqrt{-g} \; E^\mu) = 0$. 

For Klein-Gordon scalars the contribution by a single mode to the 3-vector part of the stress-energy flux is 
\be
  \cE_\mu := \langle \omega l m | E_\mu | \omega l  m \rangle = -\langle \omega l m | T_{\mu t} | \omega l m \rangle \,,
\ee
where $T_{\mu\nu} = \partial_\mu \hat\psi \, \partial_\nu \hat \psi - \frac12 g_{\mu\nu} (\partial \hat\psi)^2$. Here $\hat \psi (x) = \sum_\ssA [\psi_\ssA(x) \hat{a}_\ssA + \psi^*_\ssA(x) \hat{a}^\dagger_\ssA]$ denotes the field operator, where the mode labels $A = \{\omega, l, m\}$ are temporarily restored to the mode functions. The spatial components of this vector are therefore given in terms of the mode functions by
\be
\label{Ekdef}
 \cE_k := -  \partial_t \psi \, \partial_k \psi^* - \partial_k \psi \, \partial_t \psi^*   \,.
\ee
Evaluated on energy-eigenstate modes --- those for which $i\partial_t \psi = \omega \, \psi$ --- the vectors $\cW_k$ and $\cE_k$ are not independent, since 
\be
\label{EktoWk}
  \cE_k = i \omega\,  \Bigl(  \psi \,\partial_k \psi^* - \psi^* \partial_k \psi \Bigr) = \omega \, \cW_k \,.
\ee

Klein-Gordon and \Sch\ quantities like $\cJ_k$ and $\cW_k$ are related by the expression $\psi(\bfx,t) = F_d(\bfx) \chi_d(\bfx,t)$ for real $F_d$ since we then have $\nabla \psi = F_d \nabla \chi + \chi \nabla F_d$, which when used in the definitions implies
\be
 \cW_k = F_d^2 \cJ_k \,.
\ee

The differential flux of Klein-Gordon current $\cW_k$ towards the horizon through a surface element $\exd^3 x := \exd t \, \exd \theta \, \exd \phi$ at fixed $r > r_+$ is given by
\be  
  \exd \cF := \sqrt{-\gamma} \; N_\mu \cW^\mu \, \exd^3x\,,
\ee
where $\gamma$ is the determinant of the three-dimensional metric, $\gamma_{\mu\nu}$, on this constant-$r$ surface and $N_\mu \exd x^\mu = -N_r \exd r$ is the surface element's unit normal pointing towards decreasing $r$. Using the metric \pref{Kerr} then implies
\be
 \sqrt{- \gamma} = \sqrt\Delta \; \rho \sin \theta \qquad \hbox{and} \qquad
 N_r = \frac{1}{\sqrt{g^{rr}}} = \frac{\rho}{\sqrt\Delta} \,,
\ee
and so
\be \label{Fluxvspsi}
 \exd \cF =- \Delta \sin \theta  \, \cW_r\,d^3x = -i\Delta \sin \theta  \Bigl( \psi \,\partial_r \psi^*-  \psi^* \partial_r \psi  \Bigr)\,d^3x\,.
\ee

For comparison, the differential probability flux in the flat space with $d=3$ spatial dimensions within which \pref{chieom} is the \Sch\ equation is
\be
 \exd \mathfrak{F} := -(r-r_+)^2 \sin \theta \, \cJ_r \,d^3x= -\frac{(r-r_+)^2 \sin \theta \, \cW_r}{F_3^2}\,d^3x = \frac{\exd \cF}{C_\ssF^2} \,,
\ee
(using \pref{GenFsoln} to eliminate $F_3$) which shows that $\cF$ and $\mathfrak{F}$ are proportional to one another (and are equal if $C_\ssF = 1$, as we henceforth choose). One can equally well compute the near-horizon Klein-Gordon flux directly on Kerr spacetime, or can compute the flux through the `origin' at $r-r_+ = 0$ in the corresponding \Sch\ problem.

\subsubsection{Near-horizon probability flow}

Evaluating the differential flux, $\exd \cF$, in the Kerr near-horizon limit, $0 < r-r_+ \ll r_s$, allows use of the asymptotic solutions \pref{eq:gensoln} and \pref{psiasymptotic}, giving an expression for $\cF$ in terms of the parameter $\cR$. In this regime the asymptotic radial probability flux into the black object is proportional to
\ba \label{Fluxform}
  \exd \cF &=& -i \Delta \sin\theta ( \psi \, \partial_r \psi{}^* - \psi^* \partial_r \psi ) \nn\\
   &\simeq& \frac{\xi\Delta \sin\theta}{r-r_+}  \Bigl( \psi^- + \cR  \psi^+ \Bigr)  \Bigl( \psi^- - \cR \psi^+ \Bigr)^*+ \hbox{c.c.} \qquad \hbox{(as $r \to r_+$)}\\
  &=& 2(\omega r_s r_+-am) \sin \theta \Bigl[ |\psi^- |^2 - |\cR|^2  |\psi^+|^2 \Bigr]  \qquad \hbox{(as $r \to r_+$)} \,.  \nn
\ea
Here the second line uses the asymptotic form \pref{psiasymptotic} in the form $\partial_r \psi^\pm \simeq \pm i \xi \psi^\pm/(r-r_+)$, and the last line uses both the near-horizon expansion $\Delta(r) \simeq (2r_+-r_s)(r-r_+) = \sqrt{r_s^2 - 4a^2} \;(r-r_+)$ and the Kerr expression \pref{eq:zeta-defs=0} for $\xi$. Notice for real positive $\omega$ the flux $\exd \cF$ receives positive contributions from the in-falling mode $\psi^-$ and negative contributions from the out-going mode $\psi^+$, as expected.

Eq.~\pref{Fluxform} provides the desired connection with the reflection probability $| \cR|^2$, since it has a simple physical interpretation. By construction, the integral of $\exd \cF$ over all angles and over a time interval $T_2 - T_1$ gives the total change in integrated probability density over the entire region $r > r_+$ within this time window. Eq.~\pref{Fluxform} shows this change to be the sum of conditional probability changes: the loss, $\exd\cF^-$, of probability into the black object given that the state is $\psi_-$, plus the increase, $\exd \cF^+$, of probability coming out of the would-be horizon given that the state is $\psi^+$. $|\cR|^2$ represents the relative probability that the new physics near the would-be horizon prepares the state in $\psi^+$ rather than $\psi^-$, and as such is the reflection probability we seek.  We return in \S\ref{sec:Pheno} to a summary of how $\cR$ can be extracted from observations. 

\subsection{Boundary conditions and EFTs}
\label{ssec:BCEFTs}

How is $\cR$ related to the properties of whatever new physics intervenes near the black object's would-be horizon? Given the limitations of our understanding of this low-energy physics, rather than computing $\cR$ separately for all possible models of this physics it is more efficient to identify what features of such physics are likely to dominate at low energies. This type of question suggests using EFT methods, adapted to describe phenomena very near the horizon.   

Motivated by the equivalent \Sch\ problem we cast this EFT in terms of a boundary action defined at a surface at fixed $r$ just outside the horizon: $r = r_+ + \epsilon$ for $0 < \epsilon \ll r_s$. The idea is the requirement that the full action (`bulk' plus boundary),
\be \label{TotalS}
  S[\psi] = \int \exd^4x \sqrt{-g} \; \cL_\ssB[ \psi, \partial \psi, \cdots ] + \oint_{r = r_++\epsilon} \exd^3x \sqrt{-\gamma} \; \mathfrak{L}_\epsilon[\psi, \partial \psi, \cdots] \,,
\ee
be stationary with respect to variations $\delta\psi$ on the boundary is what generates the boundary condition at $x = r-r_+ = \epsilon$, and it is this boundary condition that determines $\cR$. Here (as above) $\gamma = \hbox{det}\, \gamma_{\mu\nu}$ is the invariant measure built from the induced metric on the surface defined by $r = r_+ + \epsilon$.  An important consistency condition requires that physical quantities (like $\cR$) not depend on the precise value of the arbitrary radius $\epsilon$ where this boundary condition is imposed, and this dictates the $\epsilon$-dependence of all of the effective couplings appearing within the EFT. 

What is important is that for low-energy applications both $\cL_\ssB$ and $\mathfrak{L}_\epsilon$ are dominated by the terms involving the fewest fields and derivatives, because on dimensional grounds the coefficients of higher powers of fields and derivatives are suppressed by higher powers of the microscopic length scale, $L \ll r_+$, associated with whatever new physics intercedes to modify General Relativity near the would-be horizon. 

In the following, we derive a boundary condition for the Klein-Gordon case $(s=0)$. In this case, the leading term of the bulk action for $\psi$ can be written in the simple form \pref{SB}~\footnote{We assume for simplicity a symmetry $\psi \to - \psi$ that forbids odd powers of $\psi$.}, plus possibly renormalizable interactions like $\mu^2 \psi^2$ and $\lambda \psi^4$, with higher-dimensional interactions like $c_6 \psi^2 (\partial \psi)^2$ or $c_8 (\partial \psi)^4$  being down by two or four powers of $L$, respectively.

For the boundary the lowest-dimension term is $\psi^2$ with no derivatives, so an interaction 
\begin{equation}\label{SP}
 \mathfrak{L}_\epsilon =  -\frac{h(\theta)}2 \, \psi^2   \,,
\end{equation}
should dominate at sufficiently low energies. $h(\theta)$ is an effective coupling, which we allow to depend on $\theta$ because the black object is only symmetric under shifts of $t$ and $\phi$. $h$ can also depend on $\epsilon$, in a way that is deduced below. 

Varying \pref{TotalS} with respect to $\psi$ --- using \pref{SB} and \pref{SP} --- then gives the boundary variation
\be
 - \oint \exd^3 x \sqrt{-\gamma}\; \Bigl[ N_\mu \partial^\mu \psi + h \, \psi \Bigr] \delta \psi = 0 
\ee
showing that it is linear `Robin' boundary conditions, $(N\cdot \partial \psi - h\, \psi)_{r=r_+ + \epsilon} = 0$, that tend to dominate at low energies. Evaluating this expression using the Kerr metric \pref{Kerr}, along the lines given above for $\exd \cF$, implies
\be\label{GenLinBC}
 \partial_r \ln \psi \Bigr|_{r-r_+=\epsilon} 
 = \left[ \frac{\rho}{\sqrt\Delta } \; h (\theta ) \right]_{r-r_+=\epsilon}
 \simeq   \sqrt{\frac{ r_+r_s - a^2 \sin^2 \theta}{(2r_+ - r_s)\epsilon}} \; h (\theta )   \,.
\ee

The boundary condition \pref{GenLinBC} is only consistent with a separated form, $\psi = R(r) S(\theta) e^{-i\omega t + im\phi}$, when the right-hand-side is $\theta$-independent, which requires 
\be \label{Simpleh}
  h(\theta) = \frac{h_0}{\rho_\epsilon(\theta)} \,,
\ee
for dimensionless constant $h_0$, where $\rho_\epsilon^2(\theta) = (r_++\epsilon)^2 + a^2 \cos^2 \theta = r_+(r_s+2\epsilon) - a^2 \sin^2 \theta +\cO(\epsilon^2)$. For other choices for $h(\theta)$ the boundary condition instead mixes modes having different quantum numbers, 
\be
  \left. \frac{\exd R_{lm}}{\exd r} \right|_{r-r_+=\epsilon} = \sum_{l'=0}^\infty C_{ll'} R_{l'm}(r_++\epsilon) \,.
\ee
for a calculable matrix $C_{ll'}$. In what follows we focus for simplicity on the case \pref{Simpleh}, which in particular includes the main case of echoes studied so far in the literature --- the spherically symmetric Schwarzschild example --- in the limit $a \to 0$. This leads us to study the boundary condition
\be\label{SimpLinBC}
  \frac{\exd \ln R}{\exd r} \Bigr|_{r-r_+=\epsilon}
 = \left[ \frac{h_0}{\sqrt\Delta } \right]_{r-r_+=\epsilon}
 \simeq \frac{h_0}{\sqrt{(2r_+ - r_s)\epsilon}} =: \lambda(\epsilon)  \,,
\ee
where the last equality defines the parameter $\lambda$.

Let us return to general spin $s$. Taking inspiration from the Klein-Gordon case $(s=0)$ we assume linear Robin boundary conditions of the form \eqref{SimpLinBC} are also valid for general spin. This is motivated by the fact that the Teukolsky equation for general spin is for a scalar quantity $\psi$, see Table~\ref{psiNP_table}, and the lowest dimensional operator we expect to be able to occur in an effective point particle action is $\bar \psi \psi$.

\subsubsection{\Sch\ formulation}

We next ask what the analog boundary condition is in the \Sch\ formulation corresponding to \pref{SimpLinBC}. If $R(r) = F_d(r) \chi_d(r)$ with $F_d$ given by \pref{GenFsoln}, then \pref{SimpLinBC} implies $\chi(r)$ satisfies
\ba
 \left.  \frac{\exd \ln\chi_d}{\exd r} \right|_{r-r_+=\epsilon} &=& - \frac{d-1}{2\epsilon}  + \left[ \frac{(s+1)\Delta'}{2\Delta} + \frac{h_0}{\sqrt\Delta } \right]_{r-r_+=\epsilon} \nn\\
 &\simeq& - \frac{d-s-2}{2\epsilon}  + \frac{h_0}{\sqrt{(2r_+ - r_s)\epsilon}}  \,,
\ea
where the approximate equality drops terms that are nonsingular as $\epsilon \to 0$. Defining \cite{PPEFT1}
\begin{equation}\label{lambdaKerr}
 \lambda_d(\epsilon) :=  \Omega_{d-1} \epsilon^{d-1} \left\{ - \frac{d-1}{2\epsilon}  + \left[ \frac{(s+1)\Delta'}{2\Delta} + \frac{h_0}{\sqrt\Delta } \right]_{r-r_+=\epsilon} \right\}  \,,
\end{equation}
where $\Omega_{d-1}$ is the volume of the unit $(d-1)$-sphere --- {\it i.e.}~$\Omega_2 = 4\pi$, $\Omega_1 = 2\pi$ and $\Omega_0 := 2$ --- then the near-horizon boundary condition becomes\footnote{The subscript `$d$' here on $\lambda_d$ is meant to distinguish the Schr\"odinger-side quantity from its analog $\lambda$ defined on the Klein-Gordon side by eq.~\pref{SimpLinBC}. }
\begin{equation}\label{chidBC}
 \Omega_{d-1} \epsilon^{d-1} \left[ \frac{\exd \ln \chi_d }{\exd r} \right]_{r=r_+ + \epsilon} = \lambda_d(\epsilon)\,,
\end{equation}
which is of the form discussed for Schr\"odinger systems in \cite{PPEFT1}. Explicitly, for $d = 1,2,3$ the boundary condition becomes:
\ba\label{chi321BC}
 2 \left[ \frac{\exd \ln \chi_1 }{\exd r} \right]_{r=r_+ + \epsilon} &=& \lambda_1(\epsilon)  \simeq  \frac{s+1}{\epsilon}  + \frac{2h_0}{\sqrt{(2r_+ - r_s)\epsilon}}  \nn\\
 2 \pi \epsilon \left[ \frac{\exd \ln \chi_2 }{\exd r} \right]_{r=r_+ + \epsilon} &=& \lambda_2(\epsilon) \simeq \frac{s}{\epsilon} +  2\pi h_0 \sqrt{ \frac{\epsilon}{2r_+ - r_s}}  \\
 4 \pi \epsilon^2 \left[ \frac{\exd \ln \chi_3 }{\exd r} \right]_{r=r_+ + \epsilon} &=& \lambda_3(\epsilon)  \simeq 2 \pi (s-1) \epsilon  + \frac{4\pi h_0 \epsilon^{3/2}}{\sqrt{2r_+ - r_s}}  \,.\nn
\ea
In the $a,s \to 0$ special case of the Schwarzschild limit we have $r_+ \to r_s$ and $\Delta \to r(r-r_s)$ and so \eqref{lambdaKerr} takes the simpler form
\begin{equation}\label{lambdaSchwz}
 \lambda_d(\epsilon) \simeq  \Omega_{d-1} \epsilon^{d-1} \left[ - \frac{d-2}{2\epsilon}  +  \frac{h_0}{\sqrt{r_s\epsilon}}   \right]  \qquad \hbox{(scalar Schwarzschild limit)} \,.
\end{equation}

There are two complementary ways to read eqs.~\pref{SimpLinBC} or \pref{chidBC}. The direct way is to regard $h_0$ as given and then solve \pref{chidBC} for the integration constant $\cR$ of \pref{eq:gensoln}. Once $\cR(h_0)$ is determined in this way physical predictions for observables (which can be computed as functions of $\cR$) are found in terms of $h_0$. In this reading $h_0$ is imagined to be computed using $\exd R/\exd r(r = r_++\epsilon)$ within whatever the full theory of new physics is of interest.

\subsection{Effective-coupling RG flow}
\label{ssec:RGFlow}

The other way to read eqs.~\pref{SimpLinBC} and \pref{chidBC} is to regard the profile $R(r)$ or $\chi_d(r)$ to be given as a function of $r$ and $\cR$, and then regard them as stating how $\lambda_d$ (and so, equivalently how $h_0$) must depend on $\epsilon$ in order for physical predictions to be $\epsilon$-independent. For all practical purposes the $\epsilon$-independence of observables is simply equivalent to demanding $\cR$ be $\epsilon$-independent. It doesn't matter for these purposes whether $\psi$ and \pref{SimpLinBC} or $\chi_d$ and \pref{chidBC} is used to determine $h_0(\epsilon)$.

This implied $\epsilon$-dependence of $h_0$ can be regarded as a renormalization group (RG) flow. As is often the case for such flows, it is useful to express it in differential form. The differential flow for $\lambda_d$ is found by using in \pref{SimpLinBC} the explicit near-horizon asymptotic forms \pref{psiasymptotic} for $R^\pm(r)$; explicitly differentiating the result with respect to $\epsilon$, holding $\cR$ fixed; and then using \pref{SimpLinBC} again to eliminate $\cR$ from the result in favour of $\lambda_d$.

\begin{figure} 
\centering
  \begin{subfloat}[][$\xi=\abs{\xi}$]%
   {    \includegraphics[width=0.45\linewidth]{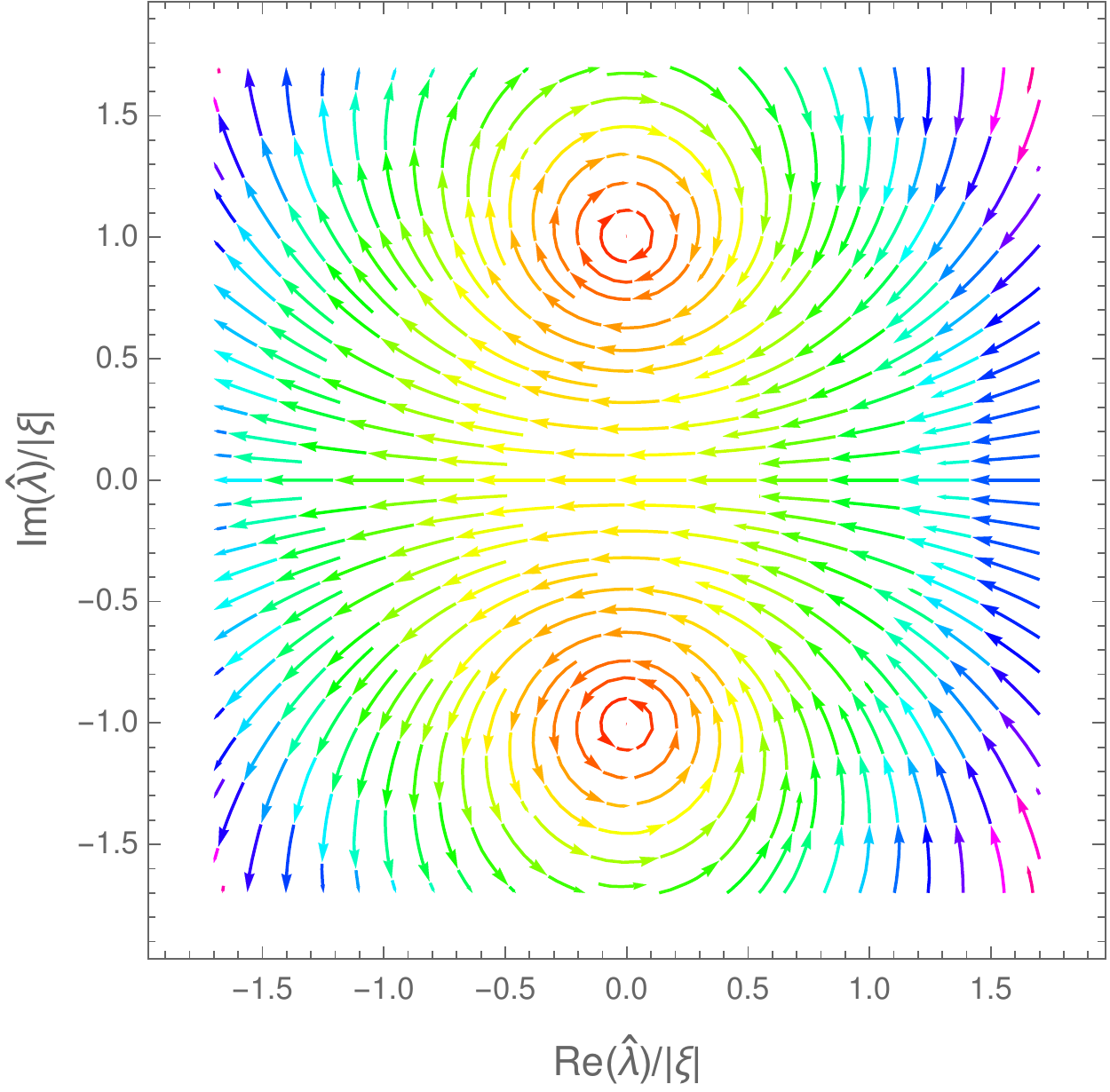}~~~
      \label{fig:phase-port-a}} %
  \end{subfloat}%
  \begin{subfloat}[][$\xi=\abs{\xi}\rm{e}^{i \pi/8}$ ]% 
   {    \includegraphics[width=0.45\linewidth]{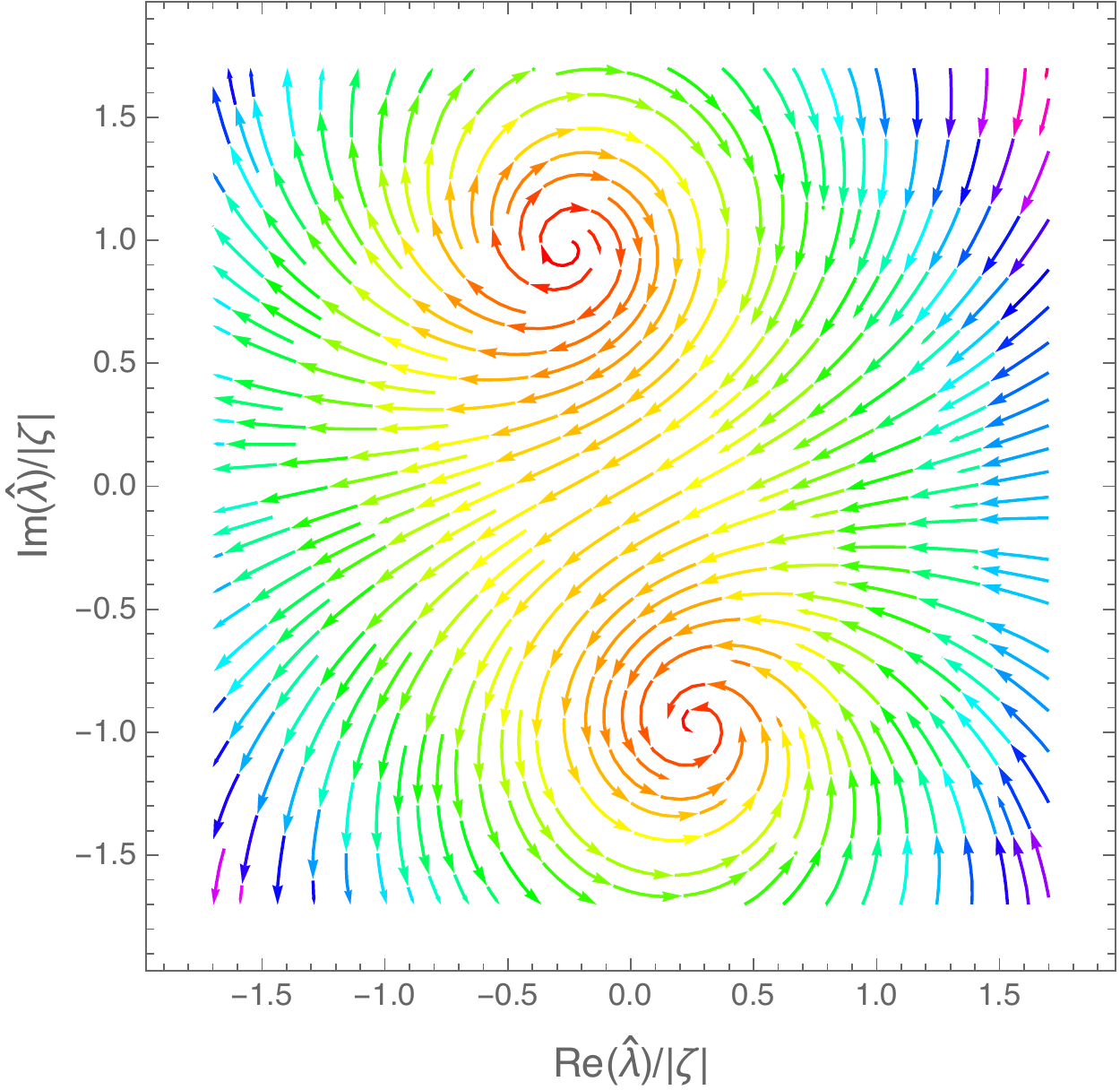}
      \label{fig:phase-port-b}} %
 \end{subfloat} 
\caption{Phase portraits of the RG flow predicted by eq.~\pref{running} and \pref{RGintegrated} for nonzero complex $\xi$. The left panel shows evolution when $\xi$ is imaginary (as appropriate, for black holes, when $\omega$ is real and $s=0$), while the right panel shows the more general case where $\xi$ has both real and imaginary parts. Colour indicates the speed with which the flow occurs with red being slow and purple being fast.}\label{fig:RG-flow}
\end{figure}

As discussed in detail in \cite{PPEFT1,PPEFT2,PPEFT3,PPEFT4}, as applied to $\chi_d$ and \pref{chidBC} these steps translate the boundary condition \pref{chidBC} into a differential evolution
\begin{equation}\label{running}
 \epsilon \frac{\exd \hat\lambda_d}{\exd\epsilon} 
 = -\frac12 \left[4\xi^2+\hat\lambda_d^2 \right]\,,
\end{equation}
with $\xi$ given in \eqref{eq:zeta-def} and
\begin{equation} \label{lambdavshat}
 \hat \lambda_d := \frac{2\lambda_d}{\Omega_{d-1} \epsilon^{d-2}} + (d-2)\,.
\end{equation}
The general solution to \pref{running} with initial condition $\hat \lambda_d(\epsilon_0) = \hat \lambda_{d0}$ is 
\be \label{RGintegrated}
 \frac{\hat \lambda_d(\epsilon)}{2i\xi} = \frac{(\hat \lambda_{d0} + 2i\xi)(\epsilon/\epsilon_0)^{2i\xi} + (\hat \lambda_{d0} - 2i\xi)}{(\hat \lambda_{d0} + 2 i \xi)(\epsilon/\epsilon_0)^{2i\xi} - (\hat \lambda_{d0} - 2i\xi)} \,,
\ee
in (unsurprising) agreement with \pref{chidBC} once the asymptotic near-horizon form is used for $\chi_d^\pm$. Fig.~\ref{fig:RG-flow} illustrates a flow pattern obtained in the complex $\hat\lambda_d$-plane obtained from this expression, for both real and complex $\xi$. Notice that this predicts a flow from the UV fixed point $\hat \lambda_d = + 2i\xi$ to the IR fixed point $\hat\lambda_d = - 2i\xi$.

\subsubsection{Fixed points as perfect absorbers/emitters}

The two fixed points of \eqref{running} are given by
\begin{equation}
 \hat \lambda_d = \pm 2i\xi \,.
\end{equation}
Notice that for the special case $\xi$ real, as is the case for Klein-Gordon scattering states for which $\omega$ is real, away from these fixed points the general flow is in cycles and does {\it not} run between the two fixed points.\footnote{Flow between IR and UV fixed points is generic whenever the imaginary part of $\xi$ is nonzero, such as occurs in particular if the coefficient $v_1$ of the small-$x$ inverse-square potential were such that $\sqrt{1-4v_1}$ had been real.}  This cyclic form is related to a residual discrete scale-invariance of the inverse-square potential, and also occurs in other applications, such as to Efimov states \cite{Efimov}. For complex $\xi$, as appropriate for $s\neq0$ and/or complex quasi-normal mode frequencies $\omega \ne \omega^*$, the flow instead runs from $-2i\xi$ to $+2i\xi$ as $\epsilon$ runs from 0 to $\infty$. 

Having $\hat \lambda_d$ sit at a fixed point implies --- from eq.~\pref{lambdavshat} --- the quantity $\lambda_d$ evolves as
\begin{equation} \label{lambdavshatfp}
 \lambda_d(\epsilon) = \left[ - \frac12 \, (d-2)  \pm i\xi \right]  \Omega_{d-1} \epsilon^{d-2} \qquad \hbox{(fixed point)}\,,
\end{equation}
and so from \pref{lambdaKerr} $h_0(\epsilon)$ at the fixed point evolves as
\be
 \frac{h_0 \epsilon}{\sqrt{(2r_+ - r_s)\epsilon}} = \pm i \xi - \frac{s}{2}\qquad \hbox{(fixed point)}\,.
\ee
In the scalar Schwarzschild limit this reduces to
\begin{equation}
 \lim_{a \to 0} h(\theta) = \lim_{a\to 0} \frac{h_0 (\epsilon)}{r_s}  =  \pm i \omega  \sqrt{ \frac{r_s}{\epsilon}}  \qquad \hbox{(fixed point)} \,.
\end{equation}

To gain some intuition, let us focus on the Klein-Gordon case $s=0$ for a moment. Using \pref{eq:zeta-defs=0} to eliminate $\xi$,
\be
 h(\theta) = \frac{h_0 (\epsilon)}{\rho_\epsilon} = \pm \frac{i}{\rho_\epsilon}  \left[ \frac{\omega r_s r_+-am}{\sqrt{(2r_+ - r_s)\epsilon}} \right] \qquad \hbox{(fixed point)}\,.
\ee
Notice $h_0$ is pure imaginary at the fixed point, as is known to describe more prosaic systems that can lose probability at the origin \cite{PPEFT4}. The connection between probability loss and $h$ follows if the boundary condition $N^r \partial_r \psi = (\sqrt\Delta/\rho) \partial_r \psi = h \psi$ at $r = r_++\epsilon$ is used in the definition \pref{Fluxvspsi} of probability flux, leading to
\ba
 \exd \cF (r_++\epsilon) &=&  \left. -i\Delta \sin \theta  \Bigl( \psi \,\partial_r \psi^*-  \psi^* \partial_r \psi  \Bigr) \right|_{r_++\epsilon} \nn\\
 &=&  \left. -i \sqrt\Delta\; \rho \sin \theta (h^* - h)  \psi^* \psi  \right|_{r_++\epsilon} \,.
\ea
Recall that $\exd \cF > 0$ corresponds to probability loss into the black object, corresponding to Im $h < 0$, and conversely Im $h > 0$ corresponds to outflow of probability from the black object. 

Using \pref{lambdavshatfp} in \pref{chidBC} provides a physical interpretation for what it means for the system to sit at the fixed point. Making this substitution implies
\be 
 \left. \frac{\exd \chi_d}{\exd r} \right|_{r = r_++\epsilon} =  \left.  \frac{-(d-2) \pm 2i \xi}{2(r-r_+)} \right|_{r=r_++\epsilon} \qquad \hbox{(fixed point)}\,.
\ee
which is the boundary condition satisfied by the perfectly in-falling and perfectly-outgoing solutions, $\chi_d^\pm$, corresponding to $R^\pm(r) \sim (r-r_+)^{\pm i \xi - s/2}$ in the near-horizon limit. 

These observations show that, in particular, standard black-hole perfect-absorber boundary conditions correspond in this language to the system sitting at a fixed point $\hat \lambda_d = -2i\xi$, for which $\cR = 0$. In the Schwarzschild limit this choice gives Im $h < 0$ for all $\omega > 0$, as expected. Notice that for Kerr the flux $\exd \cF$ can have either sign at the perfect-absorber fixed point, depending on the sign of the combination $\omega r_+r_s - a m$, with negative values of this combination producing probability outflow at the horizon, as appropriate for super-radiance. 

The significance of this being a fixed point is that the UV-sensitive effective coupling $\hat \lambda_d$ does not vary at all as $\epsilon \to 0$, and so as the horizon is approached. This is in accord with the usual relativistic picture for which the horizon is not locally a special place. The same is also true for the perfect emitter, which physically corresponds to a white hole; the time-reverse of a black hole. No other choices for $\hat \lambda_d$ are fixed points of the RG flow in this way and this shows that they all experience nontrivial evolution in the neighbourhood of the horizon, as might be expected if new physics begins to appear there.

\subsubsection{RG invariants and observables}

With the above formalism in place it becomes possible to specify precisely how the observable quantity $\cR$ depends on different choices for UV near-horizon physics. As argued above, this UV physics is characterized at low energies by the dimensionless effective coupling $h_0(\epsilon)$, or equivalently $\hat\lambda_d(\epsilon)$ defined in terms of $h_0$ by \pref{lambdaKerr} and \pref{lambdavshat}. 

For general $\hat \lambda(\epsilon)$ and $\epsilon$, in the case of an inverse-square potential the near-horizon reflection coefficient $\cR$ works out to be given by the formula\footnote{This expression is derived in detail in ref.~\cite{PPEFT4}, though beware that $\xi$ as used here is half as large as $\xi$ used in \cite{PPEFT4}.}  
\be \label{Rvsxilambdainvsq}
   \cR_{\text{inv-sq}} = \frac{2i\xi + \hat\lambda(\epsilon)}{2i\xi - \hat\lambda(\epsilon)}(2i \tilde k\epsilon)^{-2i\xi} \,.
\ee
In this expression $\tilde k$ is given in \eqref{ktilde} while $\xi$ is given for Kerr geometries by \pref{eq:zeta-def}, reducing for Klein-Gordon perturbations in Schwarzschild to $\xi \to \omega r_s$. 

Now for the \Sch\ analog of the Klein-Gordon field on a black-hole background, the potential of interest is {\it not} purely inverse-square, it is only asymptotically inverse-square in the near-horizon limit. In this more general situation determination of $\cR$ depends on the normalization constants --- {\it i.e.}~$N^\pm$ or $\cN^\pm$ of \pref{psiasymptotic} or \pref{chiasymptotic} --- appearing in the asymptotic form, whose determination requires solving the radial equation in the presence of the full potential. This modifies \pref{Rvsxilambdainvsq} into $\cR = \cR_{\text{UV}} \, \cR_\ssN$ where
\begin{equation}
 \cR_\ssN = \frac{N^+}{N^-}\,\left(2i\tilde k r_s\right)^{-2i \xi}\,,
\end{equation}
depends on the precise values of $N^\pm$ and $\tilde k$ but is $\epsilon$-independent, while 
\be \label{Rvsxilambda}
   \cR_{\text{UV}} = \frac{2i\xi + \hat\lambda(\epsilon)}{2i\xi - \hat\lambda(\epsilon)}\left(\frac{\epsilon}{r_s}\right)^{-2i\xi} \,.
\ee
Since our interest in what follows is specifically in the $\epsilon$-dependence of results, we put aside $\cR_\ssN$ for the moment and focus attention on $\cR_{\text{UV}}$.

An important feature about \pref{Rvsxilambda} is its RG-invariance. That is, by construction it does not depend on the precise value for $r = \epsilon$ where the boundary condition is imposed because the evolution for $\hat\lambda(\epsilon)$ given in \pref{RGintegrated} ensures it is equally true for \emph{any} $\epsilon$ located asymptotically near the horizon. More generally, this is true for any physical observable.

This RG-invariance implies that the dependence of quantities like $\cR_{\text{UV}}$ on UV physics can be more efficiently expressed using an RG-invariant characterization of the effective-coupling flow than by specifying specific values for $\epsilon$ and $h_0(\epsilon)$. A convenient choice for RG-invariant characterization of the flow defined by the RG equation \pref{running} is\footnote{This generalizes the definition of \cite{PPEFT1,PPEFT4} to the case where $\xi$ is neither real or pure imaginary.} the pair $\epsilon_\star$ and $z_\star$, with $\epsilon_\star$ defined as the value of $\epsilon$ where a particular flow line satisfies Im $[\hat\lambda_d(\epsilon_\star)/2\xi] = 0$ and $z_\star = \hbox{Re}\, [\hat\lambda_d(\epsilon_\star)/2\xi]$ defined at this point. $\epsilon_\star$ defined this way is unique for complex $\xi=\xi_{\ssR} + i \xi_{\ssI}$, since given an initial condition $\hat\lambda_d(\epsilon_0) = \hat\lambda_{d0}$ the evolution \pref{RGintegrated} implies 
\begin{equation}
 {\rm Re}\left( \frac{\hat\lambda_d}{2i\xi} \right) = {\rm Im}\left( \frac{\hat\lambda_d}{2\xi} \right) \propto  \left| \frac{\hat \lambda_{d0}}{2i \xi} + 1 \right|^2 \left( \frac{\epsilon}{\epsilon_0} \right)^{-2\xi_\ssI} -  \left| \frac{\hat \lambda_{d0}}{2i \xi} - 1 \right|^2 \left( \frac{\epsilon}{\epsilon_0} \right)^{2\xi_\ssI}  \,,
\end{equation}
and so $\rm{Im}[ \hat\lambda_d(\epsilon_\star)/2\xi] =0$ implies
\begin{equation}
  \left( \frac{\epsilon_\star}{\epsilon_0} \right)^{4\xi_\ssI} = \left|  \frac{(\hat\lambda_{d0}/2i\xi) + 1}{(\hat\lambda_{d0}/2i\xi) - 1} \right|^2 \,,
\end{equation}
which has a unique well-defined positive solution for any initial condition $\hat\lambda_{d0} \ne \pm 2i \xi$ not initially chosen at one of the fixed points. Notice that choosing the black-hole fixed point, $\hat\lambda_{d0} = - 2i \xi$ at $\epsilon = \epsilon_0$, corresponds to taking the limit $\epsilon_\star/\epsilon_0 \to 0$.

The two real numbers $\epsilon_\star$ and $z_\star$ carry precisely the same information as do $\epsilon$ and the complex value of $\hat\lambda_d(\epsilon)$. Since physical quantities are RG-invariant they can always be expressed in terms of $\epsilon_\star$ and $z_\star$ rather than $\hat \lambda_d(\epsilon)$ and $\epsilon$. In particular, evaluating \pref{Rvsxilambda} at the scale $\epsilon_\star$ gives $\cR_{\text{UV}}$ in terms of the RG invariants $\epsilon_\star$ and $z_\star$:
\be \label{RforKerr}
\cR_{\text{UV}} = \frac{i + z_\star}{i - z_\star}\left(\frac{\epsilon_\star}{r_s}\right)^{2\xi_\ssI - 2i \xi_\ssR} = \left(\frac{\epsilon_\star}{r_s}\right)^{2\xi_\ssI - 2i \xi_\ssR}  \, e^{-i \Theta_\star} \, ,
\ee
where
\be \label{Theta*def}
  \tan \Theta_\star := \frac{2z_\star}{1- z_\star^2} \,.
\ee
Notice that $\cR_{\text{UV}}$ vanishes for black-hole perfect-infall boundary conditions, as may be seen from \pref{Rvsxilambda} when $\hat \lambda_d = - 2i\xi$ or from \pref{RforKerr} when $\epsilon_\star \to 0$.

\section{Phenomenological implications}
\label{sec:Pheno}

The previous sections provide a formalism that captures how new UV physics near a would-be black hole horizon could plausibly affect the observable region outside. It is argued that this effect often enters purely as a change to the near-horizon boundary conditions encountered by modes external to the hole. Furthermore, it is argued that at low energies the relevant boundary condition is dominantly linear. 

The upshot of these earlier sections\footnote{The argument made in these earlier sections is reasonably general for Schwarzschild geometries, but (for simplicity) as derived assumes no mode mixing by the boundary conditions near the horizon for Kerr black holes.}  is that there are two RG-invariant quantities that characterize the low-energy limit of any UV physics responsible for the modified boundary condition: a dimensionless parameter $z_\star$ and a length scale $\epsilon_\star$. The absence of exotic UV physics corresponds to the limit $\epsilon_\star = 0$ or $\infty$, depending on the sign of $\xi_\ssI$, in which case physics exterior to the black hole is independent of $z_\star$. 

Previous sections compute how these parameters determine the external solution's unknown integration constant, $\cR$, with the main result being $\cR_{\text{UV}}$ given by \pref{RforKerr}. Earlier sections also show how this integration constant is related to the near-horizon reflection probability at the outer Kerr horizon, $r = r_+$. What remains is to ask how $\cR$ is related to observables measurable far from the horizon, such as during gravitational wave emission during compact-object mergers. Two potential implications are briefly described in this section. For simplicity, we once again restrict to the case of scalar fluctuations $s=0$.\footnote{The Green functions approach to the transfer function discussed in the following section is less straight forward to generalize to general spin $s$.}

\subsection{Echoes in black-hole ringdown}

The most interesting potentially measurable far-field consequence of nonzero $\cR$ is the appearance of echoes in any post-merger black hole ring-down \cite{EchoLit,EchoLitPheno,Nakano,Bueno:2017hyj,Correia:2018apm,PaniTesta,Mark:2017dnq}. Connecting $\cR$ to far-field observables requires calculating how modes penetrate from the horizon out to infinity and so requires the calculation of transmission rates (grey-body factors) through the intervening potential barrier. 

The propagation of modes out to the far-field zone is done using the relevant Green's function, whose construction does not really involve near-horizon new physics apart from through the above-described boundary condition satisfied in the near-horizon limit \cite{GreensFnRefs}.

We now derive how the transfer function $\mathcal{K}$~\eqref{eq:transfer-functionintro} can be related to $\cR_\text{UV}$. Following \cite{FrolovBook,Mark:2017dnq,Bueno:2017hyj} we work in the ``tortoise'' coordinate defined via
\begin{equation}
 d r_* = \frac{r^2+a^2}{\Delta}\,dr\,.
\end{equation}
because the asymptotic forms of the wavefunction are purely sinusoidal in this coordinate. Introducing $\tpsi = (r^2+a^2)^{1/2}\,R\,$
the radial equation \eqref{Req} becomes
\begin{equation}\label{tpsieq}
 \frac{d^2 \tpsi}{dr_*^2} + \tilde V \tpsi = 0\,,
\end{equation}
where $\tilde V$ is related to the potential $W$ in \eqref{Wdef} via a transformation of variables from $r$ to $r_*$ with asymptotic form
\begin{equation}
 \tilde V = \begin{cases}
~~  \varpi^2 & r\rightarrow r_+\\ 
\vspace{-10pt}\\                                                               
~~ \omega^2 & r\rightarrow \infty\,, \end{cases}
\end{equation}
where
\begin{equation}
 \varpi = \omega - m\, \Omega^H \qquad \text{with} \qquad \Omega^H = \frac{a}{r_+r_s}\,.
\end{equation}

The asymptotic forms that are purely outgoing at the horizon ($\tpsi_{\text{in}}$) and at spatial infinity ($\tpsi_{\text{up}}$) are the two linearly independent solutions~\cite{FrolovBook}
\begin{align}\label{eq:in-mode}
\tpsi_\text{in}&\sim \begin{cases}
~~  e^{-  i  \varpi r_*} & r\rightarrow r_+\\ 
\vspace{-10pt}\\                                                               
 A_\text{out}\, e^{  i \omega r_*} + A_\text{in}\,e^{- i \omega r_*} & r\rightarrow \infty \end{cases} \\ 
\vspace{-12pt}~ \nonumber \\
\tpsi_\text{up}&\sim \begin{cases}
B_\text{out}\, e^{  i \varpi r_*} + B_\text{in}\,e^{- i \varpi r_*} & r\rightarrow r_+\\ 
\vspace{-10pt}\\                                                               
 ~~  e^{i  \omega r_*} & r\rightarrow \infty\,, \end{cases} \label{eq:up-mode} 
\end{align}
The Wronskian $W(\tpsi_1,\tpsi_2) = \tpsi_1 (\partial_{r_*} \tpsi_2) - (\partial_{r_*} \tpsi_1) \tpsi_2 $ of two linearly independent solutions has to be a constant. This implies $\varpi B_\text{out} = \omega A_\text{in}$ [via $W(\tpsi_\text{in},\tpsi_\text{up})$ being constant], $-\varpi B_\text{in} = \omega A^*_\text{out}$ [via $W(\tpsi_\text{in}^*,\tpsi_\text{up})$], and the superradiance relation
\begin{equation}
 \left(1-\frac{m \Omega^H}{\omega} \right)^{-1} |\mathcal{T}_\text{BH}|^2 = 1 - |\mathcal{R}_\text{BH}|^2\,, 
\end{equation}
with transmission and reflection coefficients $\mathcal{T}_\text{BH} = 1/B_\text{out}$ and $\mathcal{R}_\text{BH} = B_\text{in}/B_\text{out}$ via $W(\tpsi_\text{up}^*,\tpsi_\text{up})$.

The Greens function $\cG(r_*,r_*')$ where really $\cG = \cG_{\omega m}(\theta, r_*; \theta', r_*')$, although only the dependence on $r_*$ is kept explicit is defined as a solution to equation \eqref{tpsieq} with a delta function source term:
\begin{equation} \label{Greensdefeq}
 \frac{d^2}{dr_*^2} \cG(r_*,r_*') + \tilde V(r_*) \cG(r_*,r_*') = \delta(r_* - r_*')\,.
\end{equation}
In absence of any near-horizon reflection, i.e. with purely infalling waves as $r\rightarrow r_+$ the black hole Greens function is given as
\begin{equation}
\mathcal{G}_\text{BH}(r_*,r_*')= \frac{1}{W_\text{BH}}\qty[\Theta(r_*-r_*')\tpsi_\text{up}(r_*)\tpsi_\text{in}(r_*')+\Theta(r_*'-r_*)\tpsi_\text{in}(r_*)\tpsi_\text{up}(r_*')]\,  ,
\end{equation} 
in terms of the asymptotic solutions \eqref{eq:in-mode} and \eqref{eq:up-mode} and $W_\text{BH} := W(\tpsi_\text{in},\tpsi_\text{up}) = 2 i \varpi B_\text{out}$.

For a general boundary condition that is \textit{not} purely in-falling at the horizon the solution to \eqref{Greensdefeq} is
\begin{equation}
\label{eq:greens-functiongen}
\mathcal{G}(r_*,r_*')=\mathcal{G}_\text{BH}(r_*,r_*') + \mathcal{K}\left( \frac{ \tpsi_\text{up}(r_*) \tpsi_\text{up}(r_*') }{ W_\text{BH}  } \right) \,.
\end{equation}

For the present purposes it is the quantity $\mathcal{K}$ that is of most interest, since it can be identified as the transfer function between the horizon and null infinity. Its dependence on $\mathcal{R}_\text{BH}$ and $\mathcal{T}_\text{BH}$ can be determined algebraically by demanding that $\cG(r_*,r_*')$ satisfies the new-physics boundary condition at the origin. For $r_*<r_*'$ we have
\begin{equation}
\mathcal{G}(r_*,r_*')= \frac{\tpsi_\text{up}(r_*')}{W_\text{BH}}\times[ \tpsi_\text{in}(r_*) + \mathcal{K} \tpsi_\text{up}(r_*)]~,
\end{equation}
and so taking the limit that $r\rightarrow r_+$ we find 
\begin{align}
 \begin{aligned}
  \lim_{r\rightarrow r_+}\mathcal{G}(r_*,r_*') &= \frac{\tpsi_\text{up}(r_*')}{W_\text{BH}} \qty[ e^{-i \varpi r_*}  + \mathcal{K}\qty(\frac{1}{\cT_\text{BH}}e^{  i \varpi r_*} +\frac{\mathcal{R}_\text{BH}}{\mathcal{T}_\text{BH}} e^{- i \varpi r_*})]\\
  &=\frac{\tpsi_\text{up}(r_*')}{W_\text{BH}} \qty[ \left(1 + \mathcal{K}\,\frac{\mathcal{R}_\text{BH}}{\mathcal{T}_\text{BH}}  \right)e^{-i \xi \ln x}  + \frac{\mathcal{K}}{\cT_\text{BH}} \, e^{  i \xi \ln x}]\,,
 \end{aligned}
\end{align}
where we have identified $\varpi r_* \simeq \xi \ln x$ which follows from $r_* \simeq (r_+ r_s \ln x) /\sqrt{r_s^2-4 a^2}$ in the limit $r\rightarrow r_+$ and \eqref{eq:zeta-defs=0}. To satisfy the boundary condition $\psi= \psi_- + \mathcal{R}\psi_+$ of \eqref{eq:gensoln} with the asymptotic form of $\psi_\pm$ given in \eqref{psiasymptotic} requires
\begin{equation}
\mathcal{R} = \cR_\ssN \cR_{\text{UV}} = \frac{\mathcal{K}}{\mathcal{T}_\text{BH}}\times \qty[1 + \mathcal{K}\;\frac{\mathcal{R}_\text{BH}}{\mathcal{T}_\text{BH}}]^{-1}~,
\end{equation}
which when solved for $\cK$ gives~\footnote{The transfer function in \eqref{eq:transfer-function} generalizes the result of \cite{Mark:2017dnq} which is for scalar perturbations in a Schwarzschild background to scalar perturbations in Kerr. See \cite{PaniTesta} for the generalization of \cite{Mark:2017dnq} to generic bosonic perturbations in a Schwarzschild background.}
\begin{equation}
\mathcal{K}(\omega, m,l) = \frac{ \mathcal{T}_\text{BH}(\omega,m,l) \mathcal{R}(\omega,m,l)}{1-\mathcal{R}_\text{BH}(\omega,m,l)\mathcal{R}(\omega,m,l)}\, ,
\label{eq:transfer-function}
\end{equation}
where the mode labels $\{\omega,m, l\}$ are re-instated compared to~\eqref{eq:transfer-functionintro} to remind the reader that there is an independent transfer function for each mode.

This expression for $\cK$ has a simple interpretation, made manifest using the geometric series $(1-x)^{-1} =\sum_{n=0}^\infty x^n$. This reveals the transfer function to be given by 
\begin{equation}\label{UVspacing}
\mathcal{K}=\mathcal{R}\times\qty[1 + \mathcal{R}_\text{BH} \mathcal{R} + \qty(\mathcal{R}_\text{BH} \mathcal{R})^2 + \qty(\mathcal{R}_\text{BH} \mathcal{R})^3 + ...] \times \mathcal{T}_\text{BH}\, ,
\end{equation}
and so describes how a mode reflected at the horizon escapes to infinity through transmission through the external potential barrier, possibly preceded by multiple reflections between the barrier and the horizon. 

For applications to post-merger echoes this Greens' function is to be convolved with the appropriate source in order to infer the late-time $t$-dependence of the solution. As usual, the frequency part of the integral is dominated at late times by complex poles, $\omega = \omega_{nlm}$, in the complex energy plane, corresponding to the complex eigen-frequencies of quasi-normal modes (QNMs). At late times it is the least damped of these modes that eventually dominate (and for Kerr geometries this can include growing super-radiant modes that are not damped at all).  The upshot is to require the evaluation of $\cK$ at $\omega = \omega_{nlm}$, leading to complex values for $\xi$.

\subsection{Relating UV parameters to observables}

It is the multiple reflections described by \pref{UVspacing} that are responsible for the gravitational-wave echoes, and this expression for $\cK$ shows how each successive echo is accompanied in the solution by the product $\cR_\text{BH} \cR = \cR_\text{BH} \cR_\ssN \cR_{\text{UV}}$. Should such echoes be detected the magnitude and phase of the UV-sensitive near-horizon reflection coefficient $\cR_{\text{UV}}$ can be inferred, given that the normalization contribution, $\cR_\ssN$ and barrier-reflection rate, $\cR_\text{BH}$, are both calculable using ordinary black-hole physics. The two natural quantities to measure in order to do so are echo spacings and echo damping, which together determine both the modulus and phase of $\cR_{\text{UV}}$. We discuss the dependance of each of these on $z_\star$ and $\epsilon_\star$ in turn.

\subsubsection*{Echo spacing}

A simple quantity to measure, discussed elsewhere \cite{EchoLit,EchoLitPheno,Nakano,Bueno:2017hyj,Correia:2018apm,PaniTesta,Mark:2017dnq}, is the time-delay, $\Delta t$, between successive echoes. This naively measures the travel-time between position $r=r_++\epsilon$ where the modified boundary conditions are imposed and the maximum of the external potential barrier. Indeed, some early studies explore the implications of applying Dirichlet boundary conditions for Schwarzschild geometries in this way, and found that the timing between echoes is set mostly by the time take to traverse between the horizon and the maximum of the radial potential and so gives $\Delta t \sim r_s \ln(\epsilon/r_s)$. What is puzzling about this from an EFT point of view is its explicit dependence on the arbitrary scale $\epsilon$, inconsistent with RG-invariance. In this regard we differ somewhat from the perspectives of \cite{Mark:2017dnq,Bueno:2017hyj}, and more closely parallel the point of view outlined in \cite{Correia:2018apm} where $\cR(\omega)$ is treated as a general function as opposed to a function that factorizes a phase $e^{i\omega \epsilon}$.

The analysis of previous sections shows that (for both Kerr and Schwarzschild geometries) $\Delta t$ is predicted by identifying surfaces of constant phase for $\cR \cR_\text{BH} e^{-i \omega t}$. Given expression \pref{RforKerr} for $\cR_{\text{UV}}$ the phase is
\begin{equation}
 \text{arg}\, \cR_{\text{UV}} = - \Bigl[2 \xi_\ssR \ln \left( \frac{\epsilon_{\star}}{r_s} \right) + \Theta_\star \Bigr]\,,
\end{equation}
which can be written as $\text{arg}\, \cR_{\text{UV}} = \omega_\ssR \Delta t_{\text{UV}} + \phi_{\text{UV}}$, with $\omega$-independent time-shift $\Delta t_\UV$ 
\ba \label{UVdamping}
  \Delta t_\UV &=& -\frac{2r_s r_+}{\sqrt{r_s^2 - 4 a^2}}  \ln \left( \frac{\epsilon_{\star}}{r_s} \right) \,,
\ea
and phase 
\begin{equation} \label{phaseUV}
 \phi_\text{UV} = \frac{2am}{\sqrt{r_s^2 - 4 a^2}}\ln \left( \frac{\epsilon_{\star}}{r_s} \right) - \arctan \left(\frac{2z_\star}{1-z_\star^2} \right)\,,
\end{equation}
depending on the UV-sensitive quantities $\epsilon_\star$ and $z_\star$. Here $\omega_\ssR$ denotes the real part of the relevant QNM frequency, $\omega_{nlm} = \omega_\ssR -i \omega_\ssI$. The time shift in \eqref{UVdamping} depends only logarithmically on the RG-invariant scale $\epsilon_{\star}$ while the phase shift \eqref{phaseUV} also depends in an order-unity way on $z_\star$. The total time-delay is given by $\Delta t = \Delta t_\UV + \Delta t_0$, where $\Delta t_0$ is the time-delay that is not UV-sensitive induced by $\cR_\text{BH} \cR_\ssN$.

Notice the contribution to $\Delta t$ from the $\log(\epsilon_\star/r_s)$ term is always positive in the Schwarzschild limit within the generic regime for which $\epsilon_\star \ll r_s$, but can have the opposite sign for super-radiant modes in Kerr geometries.

\subsubsection*{Echo damping}

A second simple property to measure for an echo signal is the damping in amplitude between successive echoes. This depends on UV parameters differently than the time delay because it is controlled by the modulus
\begin{equation}
 D = | \cR \cR_\text{BH}| = | \cR_{\text{UV}} \cR_\ssN \cR_\text{BH}|\,,
\end{equation}
rather than its phase. The UV-sensitive part, $|\cR_{\text{UV}}|$, can be inferred once the calculated value of $|\cR_\ssN \cR_\text{BH}|$ is removed from any measured damping. 

Eq.~\pref{RforKerr} shows that for Kerr geometries this depends on UV physics purely through the parameter $\epsilon_\star$, with a predictable dependence on mode labels $\omega$ and $m$, with
\ba
 |\cR_{\text{UV}}(\omega,m)| = \left(\frac{\epsilon_\star}{r_s}\right)^{2\xi_\ssI} &=&  \exp\left[  \frac{2 \omega_\ssI r_s r_+ }{\sqrt{r_s^2 - 4 a^2}}\, \ln \left(\frac{\epsilon_\star}{r_s}\right) \right] \qquad \hbox{(Kerr)} \\
 & \to & \left(\frac{\epsilon_\star}{r_s}\right)^{2\omega_\ssI r_s} \qquad \hbox{(Schwarzschild)} \,.\nn
\ea 
This is completely independent of $z_\star$ and varies like a power of $\epsilon_\star$ with a power proportional to the imaginary part of the relevant QNM frequency (which is always negative in the Schwarzschild case). 

In particular, $|\cR_{\text{UV}}| < 1$ is small when $\epsilon_\star > r_s$ for any damped modes (for which $\omega_\ssI < 0$), but $|\cR_{\text{UV}}| > 1$ in this same limit for super-radiant modes (for which $\omega_\ssI > 0$). Having small effects arise in the limit $\epsilon_\star > r_s$ for non-super-radiant modes is a consequence of the vanishing of $|\cR_{\text{UV}}|$ as $\epsilon_\star \to \infty$ for these modes since this corresponds to standard black-hole boundary conditions.

\section{Conclusions}
\label{sec:Concl}

A number of authors have emphasized \cite{EchoLit,EchoLitPheno,Nakano,Bueno:2017hyj,Correia:2018apm,PaniTesta,Mark:2017dnq} that exotic near-horizon deviations from standard black hole physics \cite{UVproposals, ComplexBHs} could have observable implications through echoes in the ringdown phase of gravitational wave emission from compact-object mergers. Most of the explicit analyses done to this point do so for Schwarzschild geometries, rather than the Kerr geometry more directly suited to late-time ringdown after compact-object merger.

In this paper we relate the near-horizon boundary conditions responsible for observable non-black-hole signals and relate them to a near-horizon boundary component to a low-energy EFT describing physics external to the compact object. Relating boundary conditions to an action in this way allows standard EFT arguments to classify which types of near-horizon boundary conditions might dominate at low energies. We are led in this way to linear Robin boundary conditions, \pref{RobinBCs} as being dominant for a wide class of systems. 

We use this boundary condition to compute the near-horizon reflection coefficient, $\cR(\omega,m)$, as a function of black-hole mass and angular momentum as well as mode energy and angular momentum. It is the modulus and phase of this quantity that controls potentially observable features, such as the time delay, $\Delta t$, and damping, $D$, between successive echoes. We derive how these quantities are related to the parameter $\lambda$ appearing \pref{RobinBCs} as well as on the radius $\epsilon = r-r_+$ where it is applied.

We show how the arbitrary regularization parameter $\epsilon$ drops out of physical predictions once it is renormalized into an $\epsilon$-dependence that can be regarded as being implicitly in the parameter $\lambda = \lambda(\epsilon)$ within the EFT. We compute the RG-evolution of $\lambda$ and use it to identify parameters, $z_\star$ and $\epsilon_\star$, that characterize of the UV boundary conditions in an RG-invariant way. These parameters are the only properties of UV physics to which observations are likely to be sensitive, and we compute how $\cR$ depends on them. Both parameters turn out to appear in the echo time-delay, $\Delta t$, while $\epsilon_\star$ alone appears in the damping of successive echoes. An important property of the RG evolution is that $\epsilon_\star$ need {\it not} be small, even if the range of values for $\epsilon$ must be small where the UV boundary condition is actually applied. 

By mapping the radial part of the near-horizon perturbations on a Kerr spacetime into a radial \Sch\ equation it is possible to see how the black-hole problem parallels many features of the \Sch\ problem with an inverse-square potential, with near-horizon boundary conditions mapping onto boundary conditions near the origin in the \Sch\ problem. This both provides a check on the formulation and provides access to a literature for which many of the same issues are already well-explored.

\section*{Acknowledgements}
We thank Niayesh Afshordi for helpful discussions about black-hole echoes and Horacio Camblong and Carlos Ord\'o\~nez for conversations about EFT methods near Schwarzschild horizons. This work was partially supported by funds from the Natural Sciences and Engineering Research Council (NSERC) of Canada. Research at the Perimeter Institute is supported in part by the Government of Canada through NSERC and by the Province of Ontario through MRI.

\end{document}